# Nighttime Light Intensity and Child Health Outcomes in Bangladesh

Mohammad Rafiqul Islam[a], Masud Alam[a,*], Munshi Naser İbne Afzal[a], Sakila Alam[b]

## Abstract

This paper explores the effect of urbanization on child health outcomes in Bangladesh. We use nighttime light intensity as a measure of urbanization and find that the higher intensity of nighttime light, the higher the degree of urbanization, which positively affects child health outcomes. We apply a novel methodology that combines the non-parametric and panel regression approach with the Gradient Boosting Machine that originates from machine learning algorithms. Our estimates suggest that one standard deviation increase in nighttime light intensity is associated with a 1.52 rise in Z-score of weight for age. The maximum increase of weight for height and height for age score range from 5.35 to 7.18 units. We perform several robustness tests, including a wide-ranging set of controls in generalized additive models, and find robust positive relationship holds. Our findings make several contributions: first, we rationalize our empirical findings in a utility and hybrid production function showing that urbanization's marginal effect on children's nutritional outcomes is strictly positive. Second, the relationship is nonlinear and U-shaped, where low and higher light intensity leads to poorer child health outcomes, with increases being observed to be positive at all times and along the trend line. Finally, our findings are closely linked to the effective policies in reducing children's malnutrition in low- and middle-income countries. We suggest that policies favoring small towns are more effective in improving child health outcomes than expanding megacities.

---

a. Department of Economics, Shahjalal University of Science and Technology, Sylhet, Bangladesh.
b. Department of Geography, Eden Mohila College, University of Dhaka, Bangladesh.

*. Corresponding author. Email: masudalam-eco@sust.edu

## 1. Introduction

It is well-documented that rapid expansion of urbanization in low- and middle-income countries can generate superior public health outcomes, such as improved water and sanitation, a secured standard of living, and greater access to essential and quality health services. Therefore, to the extent that we have standard measures for improved child health outcomes due to the rapid growth of urbanization, we would expect children born and residing in urban areas to have the advantages of superior health inputs and access to services and, therefore, to have better nutritional outcomes as compared to children in rural areas. In relation to the growth of urbanization, this study argues that space-based nighttime light intensity data can provide significant information for mapping urbanization, and therefore variations in intensities can distinguish rural and urban development in Bangladesh. We hypothesize that the higher intensity of nighttime light, the higher is the degree of urbanization, which positively affects three measures of child health outcomes: height-for-age (HAZ), weight-for-height (WHZ), and weight-for-age (WAZ).

Empirical studies examining the impact of urbanization on child health outcomes find that urban agglomeration leads to better child health outcomes in low- and middle-income countries. For example, Paciorek et al. (2013) find that urban children are the tallest in low and middle-income countries, while rural children in these countries are the smallest at the age of five. Srinivasan et al. (2013) also find that increasing urbanization rates are associated with a lower prevalence of child malnutrition. The interest in investigating the urban impact on child nutritional outcomes gains new momentum with the advent of satellite-based nighttime light intensity data. The advantage of this data as a marker of urbanization is that unlike survey data, in which urbanization is measured as a binary variable, satellite-based light intensity data is a continuous variable and can trace and measure the standard of living more accurately. This standard measurement of urbanization has become popular in research of recent times, particularly when it comes to assessing rural-urban differences as well as measuring maternal and child health outcomes (Amare, Arndt, Abay, & Benson, 2020; Christopher, 2010; Ghosh, Anderson, Elvidge, & Sutton, 2013; Ma, Zhou, Pei, Haynie, & Fan, 2012; Mellander, Lobo, Stolaric, & Matheson, 2015).

Although a substantial volume of studies uses the nighttime light data as a proxy measure of urbanization in middle-income countries and African and other low-income countries, little attention has been devoted to child nutritional outcomes in Bangladesh.[1] Therefore, to fill this gap, this study attempts to examine the impact of urbanization on child health outcomes in Bangladesh. Naturally, the question is, why do nutritional outcomes with higher nighttime light intensities generate better child health outcomes in urban areas? We have developed an economic model to rationalize the positive relationship between urbanization and child nutritional outcomes. This model focuses on the most robust relationship that positively links higher nighttime light intensity to a better nutritional outcome for children residing in urban areas. We applied this argument to a utility and hybrid production function. Under the standard microeconomic assumptions related to the household's utility and production function and several control variables, the interior solution of our hybrid model verifies the conjecture that the marginal effect of urbanization on children's nutritional outcomes is strictly positive.

We quantify the measure of urbanization using the 'Visible Infrared Imaging Radiometer Suite (VIIRS) Day/Night Band Nighttime Lights in 2015' published by the International Food Policy Research Institute (IFPRI). Usually, investigators apply the satellite-based nighttime light image data as a daily frequency primarily collected by the Operational Linescan System (OLS) sensors of the Defense Meteorological Satellite Program (DMSP) of the United States Air Force. Our study used the VIIRS dataset instead of DMSP's data since popular DMSP nighttime light intensity data have a few limitations in measuring urbanization. First, the DMSP data are flawed by image concealing due to cloud coverage and encrypted data conversion (Gibson, Olivia, & Boe-Gibson, 2019). Second, the DMSP data cannot also be used to predict economic performances beyond cities. In the empirical implementation, we merge 'Nighttime Lights Annual Composite' data from 'VIIRS 2015' with Bangladesh's Demographic and Health Survey (DHS) by matching cluster information of rural and urban households.

The benchmark estimation of this study employs a methodology that combines both parametric and non-parametric approaches using the Gradient Boosting Machine (GBM), K-Nearest Neighbors (KNN), and Bootstrap Aggregating that originate from machine learning algorithms. Instead of ad hoc selection of control variables, we compare the analysis of variance (ANOVA), GBM, and KNN algorithms to address endogeneity issues with an automatic choice of the degree of polynomials from the additive models. Based on the benchmark estimates, we find that urbanization is a significant determinant of child nutritional outcomes and that greater urbanization leads to a healthier child after controlling for several control variables. The pooled ordinary least squares (OLS) estimates show that one standard deviation (std. dev) increase of nighttime light intensity is associated with a 1.52 rise of Z-score of weight for age. The maximum increase of weight for height and height for age range from 5.35 to 7.18. An economic interpretation of the benchmark estimates is that the maximum semi-output elasticity value is 8.42 for weight for age when converting the values into percentage changes. In both pooled OLS and cluster fixed-effect models, we find the statistical significance of second-and fourth-degree polynomials that capture the nonlinear effects of urbanization on children's nutritional outcomes.

[1] Rahman, Mohiuddin, Kafy, Sheel, & Di (2019) use night light intensity and related remote sensing spatial characteristics to classify 331 cities of Bangladesh.

In addition to the benchmark models, we also estimate the generalized additive models (GAM) to gain a further understanding of our benchmark estimates. Consistent with the benchmark estimates, the findings in the GAMs provide a robust positive relationship between nighttime light intensity and children's nutritional outcome variables. The impact remains significant and robust for the three alternative nutritional indicators: stunted, wasted, and underweight. The probability of a child being stunted, wasted, and underweight decreases with the rise of nighttime light intensities across all regression models.

The findings of this study contribute to the ongoing debate on the effective policies in reducing children's malnutrition in low- and middle-income countries. Given the empirical evidence on the effect of urbanization on reducing poverty (Christiaensen & Kanbur, 2017; Dugoua, Kennedy, & Urpelainen, 2018; Elvidge et al., 2012; Gibson, Datt, Murgai, & Ravallion, 2017), our parametric and non-parametric estimates show that the effect of child nutritional outcomes due to a change in nighttime light intensities is likely to be larger and highly elastic than those related to other socioeconomic covariates. Suppose the policy objectives include reducing poverty and increased access to improved health services for vulnerable populations; in that case, our findings suggest that an increase in small towns with better urban amenities is more effective in improving child health outcomes than the policy towards expanding megacities.

Second, we apply a novel methodology that combines the non-parametric and panel regression approach with the Gradient Boosting Machine that originates from machine learning algorithms. furthermore, we perform several robustness tests, including a wide-ranging set of socio-economic controls in generalized additive models, and find a robust positive relationship between nighttime light intensity and children's health measures. Finally, this study is related to the literature concerning the proxy measure of urbanization and its application for examining child health nutritional outcomes in Bangladesh (Angeles et al., 2019; Smith et al., 2005; Srinivasan et al., 2013). However, previous studies are primarily based on ad hoc selection of socioeconomic covariates and a dichotomous measure of rural-urban clusters. We significantly depart by applying a machine learning algorithm that selects the model's socioeconomic covariates.

The rest of the paper has been outlined as follows: Section 2 presents an overview of the theoretical model. Section 3 starts with a description of the data sources, the measurement of nutritional outcomes, and summary statistics; then, Section 4 reports the non-parametric regressions. Next, feature selection, choice of polynomials, and the benchmark regression results are dealt with in Section 5. Finally, in Section 6, we have discussed the regression analysis using a generalized additive model, and Section 7 concludes the paper.

## 2. Theoretical Foundations

We offer a theoretical framework that provides support to the empirical relationship between the proxy measure of urbanization and child health outcomes. A complete theoretical model focusing on examining the impact of urbanization on three measures of child health would be beyond the scope of our study. Thus, we, instead, follow a simplified version of households' utility and health production function approach of Behrman and Deolalikar (1988), Grossman (1972), and Omiat and Shively (2020). We assume that households maximize their utility by choosing the consumption of child health $(CH)$ and the amount of consumption of the composite commodity[2] $(Z)$.

$$\max_{CH,Z} \quad U = f(CH, Z) \qquad ; U'_{CH} >> 0; U'_{Z} > 0$$
$$with \quad P^{CH} CH + P^{Z} Z = I$$

here, $P$ is the price, and $I$ is the household's income.

[2] Household maximizes utility conditioning on the consumption of health outputs, and other observable characteristics such as schooling, parental characteristics, family background are also suggested in Strauss and Thomas (1998). For rural Bangladesh, Khan (1984) developed a model that links households' assets and health status.

Let the production of child health follow a standard production function that includes only two purchased inputs, urbanization $(UR)$ and all other inputs[3] $(M)$.

$$Q^{CH} = F(UR, M)$$

We assume households pay the price $P^{UR}$ for urban facilities (e.g., the bill they pay for night light amenities) and $P^{M}$ for all other purchased inputs that positively provide for parental and household facilities, such as education, access to TV, and newspapers. Under the assumption of convex technologies, smooth production function, and the input budget set at $P^{UR}UR + P^{M}M = W$, the standard input demand function for these two inputs are as follows:

$$UR = UR(P^{UR}, P^{M}, P^{Z}, B)$$
$$M = M(P^{UR}, P^{M}, P^{Z}, B)$$

Here, we represent the total budget constraint of households in the following way: $B = P^{UR}UR + P^{M}M + P^{Z}Z$. We can form a hybrid production function[4] by substituting input demand function for $M$ as follows:

$$Q^{CH} = F(UR, M(P^{MR}, P^{M}, P^{Z}, B))$$

The exact identification and estimation, $Q^{CH}$, depends on the functional form of $M$. Let us assume that the functional forms are known, then households observe all the necessary information exogenously, and a rich data set provides a sufficient number of variations in the $UR$ variable across the units and over time. Moreover, assume that the dataset includes control variables related to the child, parental, and household characteristics. Under the condition of interior solution and that the second-order derivatives are satisfied, the estimated effect of $UR$ on child health measures $Q^{CH}$ can be represented as follows:

$$\left.\frac{dQ^{CH}}{dUR}\right|_{controls} = \left.\frac{dF}{dUR}\right|_{controls} + \left.\frac{dF}{dM}\frac{dM}{UR}\right|_{controls}$$

[3] Rosenzweig (1986) uses all other inputs such as schooling and medical services as an aggregate input in households' human capital production function.

[4] For optimally chosen observed inputs, Liu, Mroz, and Adair (2009) and Rosenzweig and Schultz (1983) applied a hybrid production function to estimate the marginal effect of an input.

Consistent with the fundamental objective, we hypothesize that the marginal effect of $UR$ on child health outcomes $\left.\frac{dF}{dUR}\right|_{controls} > 0$. In the subsequent empirical analysis, we examine this conjecture and show whether the empirical approach can consistently estimate a robust positive relationship between urbanization and child health outcomes.

## 3. Data and Measurement of Nutritional Outcomes

### *Sources of data*

The dataset consists of six child nutritional and health outcome variables, 14 child, parental, and household attributes, as well as nighttime light intensity as a measure of urbanization. The primary source of child health outcomes and parental and household attributes are two Bangladesh Demographic and Health Surveys (BDHS), the demographic and health surveys of the years 2011 and 2014. The BDHS is a comprehensive and widely used survey dataset that collects nationally representative data pertaining to a specific population, family planning, and maternal and child health, including geo-referenced information.

The satellite-based nighttime light intensity data comes from Advancing Research on Nutrition and Agriculture (AReNA)[5] project of the International Food Policy Research Institute (IFPRI, 2020). The dataset published by the IFPRI is a compilation of nighttime light data from the 'Version 1 Visible Infrared Imaging Radiometer Suite (VIIRS) Day/Night Band Nightime Lights' and '2015 VIIRS Nightime Lights Annual Composite' with the BDHS survey information. In addition, a wide variety of geo-referenced variables on agricultural production, agroecology, climate, demography, and infrastructure are also available in AReNA's dataset that includes Bangladesh information from 2004 to the present date. The nighttime light intensity measurement unit is a digital number ranging from 0 (no light) to 63 (the highest light intensity).

We clean and merge the dataset for the BDHS of 2011 and 2014 by employing various steps. First, we utilize the advantage of 600 primary sampling units (PSUs), also referred to as 600 rural and urban clusters in the BDHS survey, as well as the information on the latitude and longitude of each cluster to match the nearest cluster's location. We calculate the distance matrix in kilometers using latitude and longitude and define the nearest cluster located within a circular area of 1.5 kilometers. Second, we construct an approximate cluster-level panel dataset using cluster location. Third, we filter out the sampling units if the clusters or nearest clusters do not contain the three child nutritional variables and necessary household information. Finally, we merge AReNA and BDHS datasets using clusters as identifiers. The final dataset consists of 8734 household observations across seven divisions of Bangladesh. The highest proportion of the household (20.23%) is from the Chittagong division, followed by Dhaka (17%), Sylhet (13.20%), Khulna, and Barisal (12.73%), Rangpur (12.32%), and Rajshahi (11.76%). Figure 1 shows spatial disparities of the mean values of nightlight intensity, the maximum value location, and the number of households from each division in the pooled sample.

[5] The database is referred to as AReNA's DHS-GIS Database, and more details are available at https://www.ifpri.org/publication/arenas-dhs-gis-database.

**Fig 1** Spatial disparities of Nighttime Light Intensity, maximum light intensity location, and the number of households included in the pooled sample from each division.

***[insert Fig.1 here]***

***Measurement of nutritional outcomes***

The nutritional status of children's three health indices used in this study is similar to the measurement unit suggested and used by the World Health Organization (WHO) for quantifying child growth. A vast amount of recent literature, including Amare et al. (2020), Choudhury, Headey, and Masters (2019), Firestone et al. (2011), Ruel et al. (2013, 2017), Smith et al. (2005), and Van de Poel et al., (2007), use average values of these three indices, height for age, weight for height, and weight for age, to measure whether a child is stunted, wasted, and underweight, respectively.

***Summary analysis***

The descriptive statistics are presented in Table 1. The mean value of light intensity for Bangladesh in the pooled sample is found to be 1.62, with a maximum of 29.94 and a minimum of 0.00023. The density function, shown in Fig 2a-2c, suggest an approximately normal distribution of nighttime light intensity but not an exact fit to a standard normal estimation. The skewness and kurtosis appear to be suggestive of accommodating the multi-modality of our regression analysis that is often necessary for sparse data (Parzen, 1962; Silverman, 1998).

**Table 1**: Summary statistics

*Panel A: Nighttime light as a measure of urbanization*

| | Pooled sample | | | | 2014 | 2011 |
|---|---|---|---|---|---|---|
| Variable | Mean | St. Dev. | Min | Max | Mean | Mean |
| Log of Nighttime light | -1.09 | 1.787 | -8.37 | 3.39 | -0.874 | -0.964 |
| Nighttime light | 1.62 | 3.54 | 0.00023 | 29.94 | 1.912 | 1.782 |
| *N* | | *8,734* | | | *4,346* | *4,388* |

*Panel B: Child nutritional outcome variables*

| | Pooled sample | | | | 2014 | 2011 |
|---|---|---|---|---|---|---|
| Variable | Mean | St. Dev. | Min | Max | Mean | Mean |
| Height for age z score | -1.6 | 1.37 | -6.0 | 5.84 | -1.388 | -1.513 |
| Weight for height z score | -0.91 | 1.17 | -5.0 | 4.71 | -0.822 | -0.872 |
| Weight for age z score | -1.54 | 1.13 | -5.83 | 4.74 | -1.351 | -1.460 |
| Child is stunted (binary) | 0.34 | 0.47 | 0.00 | 1 | 0.314 | 0.359 |
| Child id wasted (binary) | 0.132 | 0.34 | 0.00 | 1 | 0.139 | 0.149 |
| Child is underweight (binary) | 0.30 | 0.45 | 0 | 1 | 0.281 | 0.313 |
| *N* | | *8,734* | | | *4,346* | *4,388* |

*Panel C: Child, parental, and household attributes*

| | Pooled sample | | | | 2011 | 2014 |
|---|---|---|---|---|---|---|
| Variable | Mean | St. Dev. | Min | Max | Mean | Mean |
| Years of school (mother) | 3.215 | 1.522 | 0.000 | 8.00 | 3.231 | 3.237 |
| Years of school (father) | 3.490 | 1.566 | 0.000 | 8.00 | 3.516 | 3.481 |
| Age of mother at first childbirth | 18.140 | 3.285 | 11 | 46 | 18.48 | 18.353 |
| Birth order of child for mother | 2.298 | 1.515 | 1 | 14 | 1.964 | 2.059 |
| Mother's BMI | 2,120.879 | 373.003 | 1,220.0 | 4,549.0 | 2,208.4 | 2,139.3 |
| Age of child (month) | 2.029 | 1.416 | 0.000 | 4.00 | 1.955 | 2.038 |
| Child gender (binary) | 0.515 | 0.500 | 0 | 1 | 0.513 | 0.516 |
| Poorest quantile wealth index (binary) | 0.221 | 0.415 | 0 | 1 | 0.134 | 0.110 |
| Poorer quantile wealth index (binary) | 0.193 | 0.395 | 0 | 1 | 0.163 | 0.162 |
| Middle quantile wealth index (binary) | 0.191 | 0.393 | 0 | 1 | 0.209 | 0.206 |
| Richest quantile wealth index (binary) | 0.195 | 0.397 | 0 | 1 | 0.260 | 0.286 |
| Richer quantile wealth index (binary) | 0.200 | 0.400 | 0 | 1 | 0.234 | 0.236 |
| Household owns TV (binary) | 0.409 | 0.492 | 0.00 | 1.00 | 0.520 | 0.523 |
| Household has electricity (binary) | 0.602 | 0.490 | 0.00 | 1.00 | 0.698 | 0.711 |
| *N* | | *8,734* | | | *4,346* | *4,388* |

**Fig 2** Kernel Density Estimation of Nighttime Light Intensity in Bangladesh in 2011 and 2014.

***[insert Fig.2a,2b,2c here]***

The average light intensity increased by about 7.29% from 2011 to 2014 in all survey locations. On average, 81.2% of households reside in clusters with a nighttime light intensity of 1.62 and less than 1.62 as compared to 0.055% of households residing in clusters with a nighttime light intensity greater than 25. Between 2011 and 2014, the average values of height for weight, weight for height, and weight for age of children increased by 8.26%, 5.73%, and 0.616%, respectively. The average value of a child who is stunted, wasted, and underweight is 0.34, 0.132, and 0.30 in the pooled sample, respectively. On average, 34% and 30% of the children in Bangladesh are found to be stunted and underweight for the years 2011 and 2014. This percentage of malnutrition in the survey indicates potential widespread health and nutritional risk in the children population under the age of five in Bangladesh.

## 4. Non-parametric Analysis

We apply a Nadaraya-Watson estimator (Nadaraya, 1964; Watson, 1964) and standard local polynomial nonparametric regressions to determine whether urbanization shares a linear or nonlinear relationship with the nutritional outcome variables. In order to assess the relationship, we generate nonparametric plots of means and confidence intervals for each outcome variable as a continuous smoothed function of the log of nighttime light intensity (Fig 3a-3f). For the pooled sample, the relationship is nonlinear and U-shaped across the plots 3a-3d, where low and higher values lead to poorer child health outcomes, with increases being observed to be positive at all times. The height for age scores rises sharply by an std. dev of nearly 0.032 on average for clusters residing within the lower quantile of nighttime light intensity, followed by a little drop, and then a gradual rise until the light intensity is around 1.78. In Fig 3c, the regions with a higher degree of urbanization are positively associated with children's weight for age scores within the whole range of log of nighttime light intensity, which is between -3 and 1.9. The relationship between urbanization and child weight for height scores is slightly flattering for children residing in clusters that experience nighttime light intensity at a lower-medium range,

**Fig 3:** Non-parametric polynomial estimator. Relationship between Nighttime Light Intensity and child nutritional outcomes in Bangladesh. The Z-scores of nutritional outcome variables on the top panel's y-axis. The probabilities of a child being stunted, wasted, and underweight is presented on the y-axis of the bottom panel.

***[insert Fig.3a,3b,3c, 3d,3e,3f here]***

from -3.3 to -2. The relationship is significantly positive and non-linear for the clusters where the log of nighttime light intensity is greater than -2. Consistent with the findings of Amare et al. (2020), the graphs in the top panel indicate that the nighttime light intensity significantly improves children's nutritional outcomes at the lower range and after an intermediate range. The relationship is flattering for a small lower-medium range.

Once we consider the probability of a child being stunted, wasted, and underweight displayed in the bottom panel of Fig 3, the likelihood of stunting among children declines (in Fig 3d) at the lower and upper ranges. For wasted and underweight measures, Figs 3e and 3f also imply a non-linear and negative relationship across various ranges of urbanization. However, a weak relationship is observed for stunting and underweight in the lower-medium range from -4 to -3.5. Overall, non-parametric relationships in the lower-medium range indicate that the early stages have a relatively weaker influence on the child's nutritional outcomes. Conversely, the higher levels of urbanization are strongly and significantly related to better child nutritional outcomes.

## 5. Parametric Regression: Feature Selection

The non-parametric regression analysis in the previous section shows a strong positive relationship between urbanization and child nutritional outcomes. This section investigates whether this positive and significant relationship remains unchanged when regression models are controlled for child and parental characteristics. Following Smith et al. (2005), many recent studies[6] use children, parental, and household attributes such as education, nutritional status, childbirth order, age, access to household amenities, sanitary toilet facilities, pure water supply, caring practices, and family planning for children, and economic status variables to investigate the nutritional status.

[6] For example, Abdulahi et al. (2017), Akombi et al. (2017), Amare et al. (2020), and Black et al. (2013)

However, many socioeconomic, child, and parental characteristics and demographic factors appear statistically insignificant (Keino et al., 2014; Li, Kim, Vollmer, & Subramanian, 2020) in examining the prevalence and determinants of nutritional status. While adding more control variables provides a better understanding of the key variables of interest to the model, *ad hoc* selection of too many control variables also overfits the model. Three popular machine learning approaches are applied to overcome this variable selection dilemma. The machine learning algorithm compares model performance using expected test error from the models and selects the model's control variables accordingly.

We evaluate the 14 features of the child, family, and household features from panel c of Table 1 and up to fourth-degree polynomial of the nighttime light variable using the gradient boosting machine algorithm. Figures 4a–4c show the importance of covariates for three competing models, along with the proportional significance of key predictors. While the gradient boosting machine suggests a two-degree polynomial of the nighttime light intensity variable, bagging and the K-nearest neighbor algorithm suggest the inclusion of a maximum of fourth-degree polynomials for the model. In each model, the child's age and the mother's age at the time of the child's birth are the two most important predictors, and the father and mother's education, wealth, and access to electricity are the other significant control variables. The feature importance metric provides the fundamental basis for selecting control variables for our subsequent regression analysis, such as pooled OLS and cluster fixed-effect models and the generalized additive models.

**Fig 4:** Importance of covariates for three competing models and the proportional significance of key predictors. (a) Variable importance: Bootstrap aggregating (bagging), (b) Variable importance: KNN, (c) Variable importance: Gradient Boosting Machine (GBM)

***[insert Fig.4a,4b,4c here]***

### *Parametric regression: Choice of polynomials*

Due to a strong and significant non-linear pattern in the non-parametric regression analysis, we must decide on the polynomial degree to be used for our key predictor in performing parametric polynomial regression. While the machine learning algorithm delivers the best-performing predictive model, the choice of polynomials, and the selection of controls are purely statistical and related to the set of prior covariates that we feed to the model (Durrleman & Simon, 1989; Zhang et al., 2019). Therefore, the validity of the significance of covariates is also sensitive to the selection of our predefined set and requires careful implementation of non-linear modeling of covariates (Ding, Cao, & Næss, 2018). We first apply ANOVA to check the significance of nighttime light intensity up to a fifth-degree polynomial term to address this issue. Based on the ANOVA in Table 2, either a cubic or a quadratic polynomial of nighttime intensity appears to be a reasonable fit for the data, but models with lower- or higher-order terms are not justified.

**Table 2:** ANOVA to examine the significance of polynomials of nighttime light intensity.

| | Res.Df | RSS | Df | Sum of Sq | F | Pr(>F) |
|---|---|---|---|---|---|---|
| Nightlight | 8732 | 15346 | | | | |
| Nightlight square | 8731 | 15336 | 1 | 10.1809 | 5.8021 | 0.01603 ** |
| Nightlight cubic | 8730 | 15330 | 1 | 6.2709 | 3.5738 | 0.05873* |
| Nightlight quartic | 8729 | 15318 | 1 | 11.2995 | 6.4396 | 0.01118 ** |
| Nightlight quintic | 8728 | 15315 | 1 | 3.4405 | 1.9608 | 0.16147 |

*Note:* Res.DF: Residual Degrees of Freedom, RSS: Residual Sum of Squares, Df: The difference between unconstrained model degrees of freedom and constrained model degrees of freedom, Sum of Sq: the difference in RSS, F: F-values, Pr(>F): the p-value point out the level of significance.

Following Wood (2017), we further employ an automatic smoothing choice in the additive model (Hastie, 2017; Kneib, Hothorn, & Tutz, 2009; Krantz, Suppes, & Luce, 2006) to examine if the suggested degree of polynomials in the ANOVA table is an appropriate choice for our regression models. We include intercept and the coefficients of all child and parental characteristics as parametric predictors and the smooth term of the nighttime light intensity variable. We see that the appropriate transformation for nighttime light in the additive model results in a non-linear relationship with nutritional outcomes. Furthermore, the degree of smoothing label on the vertical axis of Fig 5 shows closer to four-degree freedom for the smooth, which is also an appropriate choice for the model fit as suggested by the ANOVA. The summary output in Table 3 provides the significance of the fourth degree of smooths in the plot.

**Fig 5:** Automatic smoothing choice in the additive model

***[insert Fig.5 here]***

**Table 3:** Approximate significance of smooth of nightlight intensity

| | edf | Ref.df | F | p-value |
|---|---|---|---|---|
| s(poly(Nightlight)) | 4.848 | 5.946 | 12.97 | <0.000 *** |

s= smooth, poly= Polynomials, edf: Appropriate number of the degrees of freedom for models (amount of smoothing chosen automatically), Ref.df: A modified computation of the degrees of freedom.

***Benchmark regression: Pooled OLS***

Considering the hybrid production function in Section 3 and the longitudinal regression of Amare et al. (2020), we estimate the following baseline regression model:

$$CH_{ict} = \sum_{n=1}^{4} \beta_n (\ln_nighttime_light_{ct})^n + \gamma_1 YR_{ic} + \gamma_2 X_{ict} + \gamma_3 Cluster_c + \varepsilon_{ict}$$

here, $CH_{ict}$ is the nutritional outcome for a child $i$ from the cluster $c$ and the survey round $t$. Our variable of interest is $ln_nighttime_light_{ct}$, the natural logarithmic values of nighttime light at cluster level for periods 2007 and 2011. $YR_{ic}$ is the year dummies when the child was surveyed. $X_{ic}$ represents a vector of characteristics related to the child, parental, and household attributes that affect child health. $Cluster$ represents a set of about 600 enumeration areas (EA) that may capture time-invariant differences in child health among children living in different survey clusters. $Cluster$ dummies implement enumeration area level fixed effects.

Table 4 reports the pooled OLS and the cluster fixed-effects estimates for child health $Z$-scores. Both pooled OLS and cluster fixed-effect regression models control for the child, parental, and household characteristics. Cluster fixed-effects models are estimated using cluster-level dummies. For the three measures of child health status, urbanization is a significant determinant at levels across all the regression models. In addition, significant non-linearities are observed between urbanization and nutritional status when adding higher-order polynomials of the dependent variable in the specifications.

A stronger positive association is observed in children's weight for age (columns 3 and 6) than height for age and weight for height in Table 4. For example, in the cluster fixed-effects model, the Z score of weight for a child's age is 1.515 more than a child's weight for height and height for age. Similarly, the pooled OLS regressions' estimated coefficient suggests that a one std. dev increase in nighttime light intensity shows that the response of weight for age is 1.182 units higher than the two other Z scores. The most robust response of weight for age is expected because it measures the overall effect on a child's health. Therefore, consistent with the conjecture, we show that nighttime light intensity, which we characterize as the measure of urbanization, has a strong and statistically significant influence on a child’s nutritional outcomes. The estimated coefficient of second-and fourth-degree polynomials validates the non-linear effects of urbanization and the marginal effect of $UR$ on child health outcomes, $\frac{dF}{dUR}\Big|_{controls} > 0$. However, the estimated coefficients of third-degree polynomials are not significant at a confidence level of 95%. The findings broadly support those in previous literature (Amare et al., 2020; Christiaensen, De Weerdt, & Todo, 2013; Smith et al., 2005), which also show weaker association in a higher stage of urbanization and strong influence at the early stage of urbanization.

Following Baltagi and Liu (2008) and  Millo and Piras (2012), we perform a *Lagrange Multiplier (LM)* test[7] to verify that our regression results do not suffer from spatially autocorrelated errors, and therefore, estimators are unbiased and consistent. Under the null hypothesis, no spatial lag dependence in the model’s residuals, and the specification satisfies the usual assumptions related to fixed effects panel model. We find that the null hypothesis of *no spatial lag dependence* is not rejected (LM = 5.7034, df = 3, *p-value* = 0.127). The test provides evidence of the absence of spatial autocorrelation in the fixed effects panel data specification.

[7] We also perform the Wooldridge test for autocorrelation in panel data. The null hypothesis of *No first-order autocorrelation* is not rejected at F(1, 10) = 1.67459 with *p-value* = P(F(1, 10) > 1.67459) = 0.224732.

We now provide an economic understanding of regression coefficients by calculating the percentage change of our empirical results presented in Table 4. A common economic interpretation for the computed value is elasticity. Interpreted as a semi-output elasticity[8] in the hybrid production model, the benchmark estimates in the fixed-effects model generate an elasticity value of 8.42 for weight for age and the values of 6.72 and 6.15 for weight for height and height for age, respectively. Similarly, for the pooled OLS estimates, a one percent increase in nighttime light intensity increases the Z score of child health outcomes by a maximum of 7.18 units.

Looking at the findings in Table 5, pooled OLS and cluster fixed-effects models show the binary measure of a child's nutritional status in terms of stunting, wasting, and underweight. The estimated coefficients in columns (1) through (6) indicate that the child's nutritional status improves with an increase in nighttime light intensity. The average marginal effect of a child being underweight decreases by 1.78 units due to a one std. dev increase in light intensity after controlling for child, mother, and household characteristics.

To provide the robustness of the relationship, we run a similar regression for the survey years 2011 and 2014. Tables 6 and 7 report the findings from the OLS and the cluster fixed-effects models. Overall, the estimated coefficients for children's nutritional outcome variables show strong positive and non-linear relationships between urbanization and child health outcomes. Consistent with the previous findings, our findings show that among the three child's health measures, the association with urbanization is the most substantial for weight for age. Moreover, the magnitude of estimated coefficients varies significantly across the degree of urbanization when the models control for a set of child and parental characteristics.

[8] From $CH_{ict} = \sum_{n=1}^{4} \beta_n (ln_nighttime_light)^n + \gamma Controls + \varepsilon$, we calculate elasticity by taking the unit change in $CH_{ict}$resulting from a percentage change in nighttime light intensity. Using simple calculus differential of the estimated equation provides:$d(CH_{ict}) = \beta * d(nighttime_light)/nighttime_light$. Divide both sides by 100 to get percentage, and therefore $(\beta/100)$ is the change in $CH_{ict}$ measured in units from a one percent increase in nighttime light intensity.

**Table 4:** Pooled OLS and cluster fixed effect model as a reference or baseline model (Pooled sample)

| | *Dependent variables:* | | | | | |
|---|---|---|---|---|---|---|
| | Pooled OLS | | | Cluster FE | | |
| | HAZ | WHZ | WAZ | HAZ | WHZ | WAZ |
| | (1) | (2) | (3) | (4) | (5) | (6) |
| $(\text{Nightlight})^1$ | $5.358^{***}$ | $6.103^{***}$ | $7.182^{***}$ | $6.152^{***}$ | $6.727^{***}$ | $8.242^{***}$ |
| | (1.442) | (1.313) | (1.224) | (2.113) | (1.937) | (1.793) |
| $(\text{Nightlight})^2$ | 2.017 | $2.455^{**}$ | $2.939^{***}$ | $4.777^{**}$ | 1.162 | $3.639^{**}$ |
| | (1.294) | (1.179) | (1.099) | (2.016) | (1.848) | (1.710) |
| $(\text{Nightlight})^3$ | -2.001 | 0.742 | -0.752 | -0.227 | -0.116 | -0.066 |
| | (1.296) | (1.181) | (1.100) | (1.898) | (1.740) | (1.610) |
| $(\text{Nightlight})^4$ | $-2.555^{**}$ | $-2.532^{**}$ | $-3.353^{***}$ | -0.972 | $-2.969^{*}$ | $-2.688^{*}$ |
| | (1.295) | (1.180) | (1.099) | (1.892) | (1.735) | (1.606) |
| Survey year dummy (2014=1) | $0.105^{***}$ | 0.038 | $0.090^{***}$ | $0.091^{***}$ | $0.052^{**}$ | $0.091^{***}$ |
| | (0.028) | (0.025) | (0.024) | (0.029) | (0.027) | (0.025) |
| Mother's education | $0.020^{**}$ | -0.006 | 0.006 | $0.021^{**}$ | -0.005 | 0.008 |
| | (0.009) | (0.008) | (0.008) | (0.010) | (0.009) | (0.008) |
| Age of mother at first birth | $0.049^{***}$ | $0.016^{***}$ | $0.040^{***}$ | $0.043^{***}$ | $0.016^{***}$ | $0.037^{***}$ |
| | (0.004) | (0.004) | (0.004) | (0.005) | (0.004) | (0.004) |
| Age of child | $-0.146^{***}$ | $-0.109^{***}$ | $-0.155^{***}$ | $-0.145^{***}$ | $-0.107^{***}$ | $-0.152^{***}$ |
| | (0.010) | (0.009) | (0.008) | (0.010) | (0.009) | (0.009) |
| Wealth dummy | $-0.075^{*}$ | $-0.095^{***}$ | $-0.113^{***}$ | -0.063 | $-0.084^{**}$ | $-0.096^{***}$ |
| (poorest wealth=1) | (0.040) | (0.037) | (0.034) | (0.042) | (0.039) | (0.036) |
| Electricity dummy | $0.249^{***}$ | $0.157^{***}$ | $0.260^{***}$ | $0.274^{***}$ | $0.148^{***}$ | $0.270^{***}$ |
| (have access=1) | (0.035) | (0.032) | (0.030) | (0.040) | (0.036) | (0.034) |
| Observations | 8,734 | 8,734 | 8,734 | 8,734 | 8,734 | 8,734 |
| $R^2$ | 0.064 | 0.034 | 0.088 | 0.054 | 0.028 | 0.075 |
| Adjusted $R^2$ | 0.063 | 0.033 | 0.087 | -0.016 | -0.045 | 0.005 |
| F Statistic | $59.527^{***}$ (df = 10; 8723) | $30.959^{***}$ (df = 10; 8723) | $84.546^{***}$ (df = 10; 8723) | $46.795^{***}$ (df = 10; 8124) | $23.005^{***}$ (df = 10; 8124) | $65.454^{***}$ (df = 10; 8124) |

*Note:* $^{*}p<0.1$; $^{**}p<0.05$; $^{***}p<0.01$

**Table 5:** Pooled OLS and cluster fixed effects model as a reference or baseline model (Pooled sample)

| | *Dependent variables:* | | | | | |
|---|---|---|---|---|---|---|
| | Pooled OLS | | | Cluster FE | | |
| | stunted | wasted | underweight | stunted | wasted | underweight |
| | (1) | (2) | (3) | (4) | (5) | (6) |
| $(\text{Nightlight})^1$ | -1.012** | -0.679* | -1.785*** | -1.292* | -0.847 | -1.759** |
| | (0.513) | (0.389) | (0.495) | (0.755) | (0.577) | (0.732) |
| $(\text{Nightlight})^2$ | 0.298 | 0.072 | -0.382 | -0.359 | -0.222 | -0.887 |
| | (0.464) | (0.352) | (0.448) | (0.723) | (0.552) | (0.701) |
| $(\text{Nightlight})^3$ | 0.519 | -0.070 | 0.179 | -0.190 | 0.062 | -0.477 |
| | (0.465) | (0.352) | (0.449) | (0.680) | (0.520) | (0.660) |
| $(\text{Nightlight})^4$ | 0.274 | 0.270 | 0.816* | 0.049 | 0.558 | 0.472 |
| | (0.464) | (0.352) | (0.448) | (0.679) | (0.519) | (0.658) |
| Survey year dummy (2014=1) | -0.044*** | -0.010 | -0.030*** | -0.038*** | -0.009 | -0.026*** |
| | (0.010) | (0.008) | (0.010) | (0.010) | (0.008) | (0.010) |
| Mother's education | -0.012*** | 0.003 | -0.001 | -0.012*** | 0.003 | -0.003 |
| | (0.003) | (0.002) | (0.003) | (0.003) | (0.003) | (0.003) |
| Age of mother at first birth | -0.013*** | -0.0004 | -0.009*** | -0.012*** | -0.001 | -0.009*** |
| | (0.002) | (0.001) | (0.001) | (0.002) | (0.001) | (0.002) |
| Age of child | 0.030*** | 0.004 | 0.041*** | 0.029*** | 0.004 | 0.040*** |
| | (0.004) | (0.003) | (0.003) | (0.004) | (0.003) | (0.003) |
| Wealth dummy (poorest wealth=1) | 0.135*** | 0.006 | 0.104*** | 0.135*** | 0.005 | 0.104*** |
| | (0.018) | (0.013) | (0.017) | (0.019) | (0.014) | (0.018) |
| Electricity dummy (have access=1) | -0.043*** | -0.030*** | -0.063*** | -0.047*** | -0.023** | -0.065*** |
| | (0.014) | (0.010) | (0.013) | (0.015) | (0.012) | (0.015) |
| Observations | 8,734 | 8,734 | 8,734 | 8,734 | 8,734 | 8,734 |
| $R^2$ | 0.043 | 0.004 | 0.045 | 0.036 | 0.002 | 0.038 |
| Adjusted $R^2$ | 0.041 | 0.002 | 0.044 | -0.036 | -0.072 | -0.034 |
| F Statistic | 38.762*** (df = 10; 8723) | 3.185*** (df = 10; 8723) | 41.475*** (df = 10; 8723) | 30.609*** (df = 10; 8124) | 1.930** (df = 10; 8124) | 32.056*** (df = 10; 8124) |
| *Note:* | *p<0.1; **p<0.05; ***p<0.01 | | | | | |

**Table 6:** Pooled OLS and cluster effects model as a reference or baseline model (Year 2011)

| | *Dependent variable:* | | | | | |
|---|---|---|---|---|---|---|
| | OLS | | | Cluster FE | | |
| | HAZ | WHZ | WAZ | HAZ | HAZ | HAZ |
| | (1) | (2) | (3) | (4) | (5) | (6) |
| $(Nightlight)^1$ | 3.860*** | 4.108*** | 5.043*** | 4.168*** | 4.434*** | 5.480*** |
| | (1.444) | (1.314) | (1.221) | (1.583) | (1.522) | (1.386) |
| $(Nightlight)^2$ | 2.116 | 2.398** | 3.107*** | 2.078 | 2.527* | 3.123** |
| | (1.323) | (1.203) | (1.119) | (1.458) | (1.407) | (1.280) |
| $(Nightlight)^3$ | -3.276** | 0.289 | -1.666 | -3.324** | 0.173 | -1.753 |
| | (1.321) | (1.201) | (1.117) | (1.439) | (1.379) | (1.257) |
| $(Nightlight)^4$ | 1.173 | -1.336 | -0.412 | 1.142 | -1.459 | -0.479 |
| | (1.325) | (1.205) | (1.120) | (1.441) | (1.381) | (1.259) |
| Mother's education | 0.023* | 0.001 | 0.014 | 0.024* | 0.003 | 0.016 |
| | (0.013) | (0.012) | (0.011) | (0.013) | (0.012) | (0.011) |
| Age of mother at first birth | 0.051*** | 0.016*** | 0.041*** | 0.048*** | 0.017*** | 0.040*** |
| | (0.006) | (0.006) | (0.005) | (0.006) | (0.006) | (0.005) |
| Age of child | -0.128*** | -0.121*** | -0.152*** | -0.129*** | -0.119*** | -0.151*** |
| | (0.014) | (0.013) | (0.012) | (0.014) | (0.013) | (0.012) |
| Wealth dummy (poorest wealth=1) | -0.110* | -0.078 | -0.113** | -0.097* | -0.072 | -0.099** |
| | (0.059) | (0.053) | (0.050) | (0.059) | (0.054) | (0.050) |
| Electricity dummy (have access=1) | 0.270*** | 0.148*** | 0.271*** | 0.270*** | 0.135*** | 0.265*** |
| | (0.050) | (0.046) | (0.043) | (0.052) | (0.048) | (0.044) |
| Constant | -2.428*** | -1.021*** | -2.130*** | -2.382*** | -1.030*** | -2.102*** |
| | (0.126) | (0.115) | (0.107) | (0.128) | (0.117) | (0.108) |
| Observations | 4,388 | 4,388 | 4,388 | 4,388 | 4,388 | 4,388 |
| $R^2$ | 0.060 | 0.035 | 0.086 | 0.056 | 0.033 | 0.082 |
| Adjusted $R^2$ | 0.058 | 0.033 | 0.085 | 0.055 | 0.031 | 0.080 |
| F Statistic (df = 9; 4378) | 31.092*** | 17.745*** | 46.010*** | 258.007*** | 144.666*** | 376.500*** |
| *Note:* | *p<0.1; **p<0.05; ***p<0.01 | | | | | |

**Table7:** OLS and cluster fixed effects model as a reference or baseline model (Survey year 2014)

| | *Dependent variables:* | | | | | |
|---|---|---|---|---|---|---|
| | OLS | | | Cluster FE | | |
| | HAZ | WHZ | WAZ | HAZ | WHZ | WAZ |
| | (1) | (2) | (3) | (4) | (5) | (6) |
| $(\text{Nightlight})^1$ | 3.766*** | 4.518*** | 5.158*** | 3.731** | 4.542*** | 5.240*** |
| | (1.441) | (1.316) | (1.229) | (1.680) | (1.358) | (1.410) |
| $(\text{Nightlight})^2$ | 0.755 | 1.303 | 1.231 | 1.334 | 1.296 | 1.521 |
| | (1.267) | (1.157) | (1.081) | (1.550) | (1.207) | (1.294) |
| $(\text{Nightlight})^3$ | -0.762 | 0.294 | -0.455 | -1.023 | 0.287 | -0.622 |
| | (1.270) | (1.160) | (1.084) | (1.537) | (1.207) | (1.284) |
| $(\text{Nightlight})^4$ | -2.676** | -2.133* | -3.046*** | -2.664* | -2.122* | -3.043** |
| | (1.265) | (1.156) | (1.080) | (1.487) | (1.194) | (1.246) |
| Mother's education | 0.016 | -0.013 | -0.002 | 0.014 | -0.014 | -0.003 |
| | (0.013) | (0.012) | (0.011) | (0.013) | (0.012) | (0.011) |
| Age of mother at first birth | 0.048*** | 0.017*** | 0.039*** | 0.046*** | 0.016*** | 0.037*** |
| | (0.006) | (0.005) | (0.005) | (0.006) | (0.005) | (0.005) |
| Age of child | -0.166*** | -0.097*** | -0.158*** | -0.166*** | -0.097*** | -0.158*** |
| | (0.014) | (0.013) | (0.012) | (0.014) | (0.013) | (0.012) |
| Wealth dummy (poorest wealth=1) | -0.040 | -0.112** | -0.112** | -0.036 | -0.112** | -0.107** |
| | (0.055) | (0.050) | (0.047) | (0.055) | (0.051) | (0.047) |
| Electricity dummy (have access=1) | 0.227*** | 0.164*** | 0.248*** | 0.240*** | 0.164*** | 0.255*** |
| | (0.049) | (0.045) | (0.042) | (0.052) | (0.045) | (0.044) |
| Constant | -2.162*** | -0.991*** | -1.916*** | -2.111*** | -0.984*** | -1.872*** |
| | (0.120) | (0.110) | (0.102) | (0.122) | (0.110) | (0.104) |
| Observations | 4,346 | 4,346 | 4,346 | 4,346 | 4,346 | 4,346 |
| $R^2$ | 0.067 | 0.033 | 0.087 | 0.066 | 0.032 | 0.085 |
| Adjusted $R^2$ | 0.065 | 0.031 | 0.085 | 0.064 | 0.030 | 0.083 |
| F Statistic (df = 9; 4336) | 34.324*** | 16.572*** | 45.979*** | 285.762*** | 144.948*** | 378.575*** |
| *Note:* | *p<0.1; **p<0.05; ***p<0.01 | | | | | |

## 6. Robustness Analysis: Generalized Additive Model

The advantage of non-parametric models is that the model fit avoids the need for parametric assumptions. However, fitting a non-parametric regression is impractical and often difficult when the response variable requires more than two or three control variables in the model or when there is a need for a large sample size. Similarly, a limitation of the parametric approach is that the model considers only the polynomial terms of nighttime light intensity at once. So, the functional form and estimated coefficients may not provide the best predictive accuracy, especially when the polynomials of nighttime light intensity also require smoothing other variables simultaneously.

A good compromise between a linear parametric and non-parametric regression is the use of the generalized additive models (GAMs) to overcome some of the limitations mentioned above.[9] The GAMs are an extension of generalized linear models to the family of additive models, where non-linear effects of continuous variables can be modeled more flexibly by linking smooth functions and the expected value of a response variable (Friedman, Hastie, & Tibshirani, 2001; Stone, 1985; Wood, 2017). We consider a response variable $y \in [HAZ, WHZ, WAZ, stunted, wasted, underweight]$ and the $p$ vector of continuous and categorical covariates $x_1, x_2 \cdots x_p$, then the effect of the covariates on the response variable can be expressed through the following linear predictor:

$$\eta = \beta_0 + \beta_1 x_1 + \beta_2 x_2 + \cdots + \beta_p x_p = x^T \beta$$

Here, the conditional mean response in our linear model is $Ey = \mu$. Let a link function $g(.)$ specify the connection between the mean response $\mu$ and the vector of covariates through[10] $g(\mu) = \eta = x\beta$, where $VAR(y) \propto V(\mu)$, and $V(\mu) = \mu(1-\mu)/n$ for a sample size of $n$. Following Hastie (2017) and Wood (2017), we write our GAM in the following manner:

[9] A few recent noteworthy applications of GAMs in children's health and nutritional outcomes are in Bacha (2020), Dong et al. (2019), Hebestreit (2017), and Hunter & Prüss-Ustün (2016).

[10] An obvious choice of the link function is $\eta = \mu$ for the family of Gaussian model, but an alternative selection of $y = g^{-1}(x^T\beta) + \upsilon$ is also suggested in the theoretical literature (Agresti, 2015; Fox, 2015; Nelder & Wedderburn, 1972).

$$y = \beta_0 + \sum_{j=1}^{p} S_j(x_j) + Z\varphi + \varepsilon$$

Here, $S_j(x_j)$, $j = 1,2,\cdots,k$ is the additive components that permit a simultaneous smoothing transformation without any parametric assumption; for the design matrix $Z$, the random part $Z\varphi$ is not modeled additively. The matrix $Z$ includes both continuous and categorical variables, such as child and parental characteristics.

We follow Hastie and Tibshirani (1990) and utilize functional facilities from the popular R-package GAM (Hastie & Hastie, 2020) and Mixed GAM Computation Vehicle (MGCV) (Wood, 2012). We fit the model using both non-parametric spline smoothers and the parametric polynomial regression, where we allow an automatic selection of the degree of smoothing for nighttime light intensity. The regression outputs are qualitatively similar in both parametric and non-parametric regressions. Therefore, we report parametric regression output in Tables 8 and 9.

Consistent with the baseline OLS regression, generalized additive model regression output in Table 8 shows that the nighttime light intensity is positively and significantly related to children's Z-score of nutritional outcome variables. The significant impact also remains the same for three other nutritional indicators: the decreasing probability of a child being stunted, wasted, and underweight.  The higher-order polynomial coefficients of nighttime light intensity presented in Tables 8 and 9 show a significant non-linear relationship with nutritional outcomes. While the first- and second-order polynomials are economically consistent with the theoretical analysis in Section 2, the third-order and a few fourth-degree polynomials are not statistically significant in both tables. Noticeably, estimated negative coefficients of the fourth-degree polynomial are not consistent with the empirical literature, which shows the positive effect of urban expansion on the reduction of poverty (Christiaensen & Kanbur, 2017, Gibson et al., 2017) and improvement of child well-being (Amare et al., 2020; Smith et al., 2005). In addition, the machine learning methods are likely to be overfitting the data. We prefer to leave this inconsistency to the lack of theoretical guidance in selecting a set of appropriate controls in the models and need to be improved in future research.

**Table 8:** Generalized Additive Model (pooled)

| | *Dependent variables:* | | |
|---|---|---|---|
| | WAZ | WHZ | HAZ |
| | (1) | (2) | (3) |
| $(\text{Nightlight})^1$ | 9.049*** | 7.091*** | 7.348*** |
| | (1.220) | (1.299) | (1.437) |
| $(\text{Nightlight})^2$ | 3.968*** | 2.896** | 3.246** |
| | (1.105) | (1.177) | (1.302) |
| $(\text{Nightlight})^3$ | -0.355 | 0.882 | -1.479 |
| | (1.107) | (1.179) | (1.304) |
| $(\text{Nightlight})^4$ | -3.291*** | -2.495** | -2.496* |
| | (1.108) | (1.181) | (1.306) |
| Mother's education | 0.017** | -0.001 | 0.033*** |
| | (0.008) | (0.008) | (0.009) |
| Age of child | -0.157*** | -0.110*** | -0.149*** |
| | (0.008) | (0.009) | (0.010) |
| Electricity dummy (have access=1) | 0.320*** | 0.192*** | 0.306*** |
| | (0.029) | (0.031) | (0.034) |
| Constant | -1.374*** | -0.758*** | -1.477*** |
| | (0.038) | (0.040) | (0.045) |
| Observations | 8,734 | 8,734 | 8,734 |
| Adjusted $R^2$ | 0.072 | 0.030 | 0.047 |
| Log Likelihood | -13,268.580 | -13,823.420 | -14,698.920 |
| UBRE | 1.222 | 1.387 | 1.695 |
| *Note:* | *p<0.1; **p<0.05; ***p<0.01 | | |

**Table 9:** Generalized Additive Model (pooled)

| | *Dependent variables:* | | |
|---|---|---|---|
| | stunted | wasted | underweight |
| | (1) | (2) | (3) |
| $(\text{Nightlight})^1$ | -1.488$^{***}$ | -0.710$^{*}$ | -2.113$^{***}$ |
| | (0.515) | (0.387) | (0.496) |
| $(\text{Nightlight})^2$ | -0.168 | 0.048 | -0.714 |
| | (0.467) | (0.350) | (0.449) |
| $(\text{Nightlight})^3$ | 0.426 | -0.057 | 0.115 |
| | (0.468) | (0.351) | (0.450) |
| $(\text{Nightlight})^4$ | 0.295 | 0.266 | 0.835$^{*}$ |
| | (0.468) | (0.351) | (0.450) |
| Mother's education | -0.015$^{***}$ | 0.003 | -0.004 |
| | (0.003) | (0.002) | (0.003) |
| Age of child | 0.031$^{***}$ | 0.004$^{*}$ | 0.041$^{***}$ |
| | (0.004) | (0.003) | (0.003) |
| Electricity dummy (have access=1) | -0.104$^{***}$ | -0.032$^{***}$ | -0.109$^{***}$ |
| | (0.012) | (0.009) | (0.012) |
| Constant | 0.397$^{***}$ | 0.149$^{***}$ | 0.304$^{***}$ |
| | (0.016) | (0.012) | (0.015) |
| Observations | 8,734 | 8,734 | 8,734 |
| Adjusted $R^2$ | 0.025 | 0.003 | 0.035 |
| Log Likelihood | -5,742.679 | -3,237.314 | -5,403.901 |
| UBRE | 0.218 | 0.123 | 0.202 |
| *Note:* | $^{*}$p<0.1; $^{**}$p<0.05; $^{***}$p<0.01 | | |

Finally, we perform several diagnostic tests to validate our results. We estimate the percentage of dispersion predicted from the GAMs, both with and without the smoothed component of nighttime light intensity, in order to determine the explanatory power of the fitted models. This power is measured as the additional percentage of deviance explained from the fitted models. This additional dispersion is the percentage difference of deviance explained between the model with smoothed nighttime light intensity and the unexplained dispersion of the model with no smooth nighttime light intensity variables. The transformation of the nighttime light intensity variable is strongly significant across all three nutritional outcome variables, and the model fit is relatively superior with the suggested degree of smoothness than the baseline OLS models. For height for age, weight for age, and weight for height, the additional deviance explained by the models with the smooth component are 0.05%, 0.04%, and 0.18%, respectively.

## 7. Conclusion

This paper combines parametric and non-parametric regression models with the help of machine learning algorithms to study the impact of urbanization on child health in Bangladesh. We use nighttime light intensity data as a proxy measure of urbanization. We find that the effect of urbanization is positive and robust across various measures of child health. The magnitude of the response is non-linear and that the estimated coefficients are significant for all the standard measures of child nutritional outcomes. Machine learning algorithms related to child health outcomes' responses to child and parental features suggest that parental education, child age, birth order, access to electricity, and wealth are significant determinants of this relationship. For children with parents who have had higher schooling years, a higher share of wealth, and more access to electricity, the nighttime light intensity variable seems to have a strong positive effect on their health status.

We develop a hybrid production function for an economic interpretation of the conjecture that nighttime light intensity is a powerful determinant of the development of a child's health. The results show that a one percent increase in nighttime light intensity significantly increases the Z score of child health outcomes by a maximum of 8.42 units. This positive relationship appears remarkably robust across parametric and non-parametric models and when the child and parental characteristics are considered as the control variables. A hybrid version of the health production function and household utility function provides theoretical and economic support to our empirical findings.

The results highlight the greater impact of the expansion of towns, investments in public infrastructure, and technology on the improvement of child nutritional outcomes than the expansion of megacities. We would be interested in exploring how urban features, country heterogeneities, and their nonlinear relationship can be incorporated into a theoretical model in future work. Moreover, it would also be interesting to examine, in detail, the association between the degree of urbanization and country characteristics and the role of country-specific factors in influencing the development of the health of children in low- and middle-income countries.

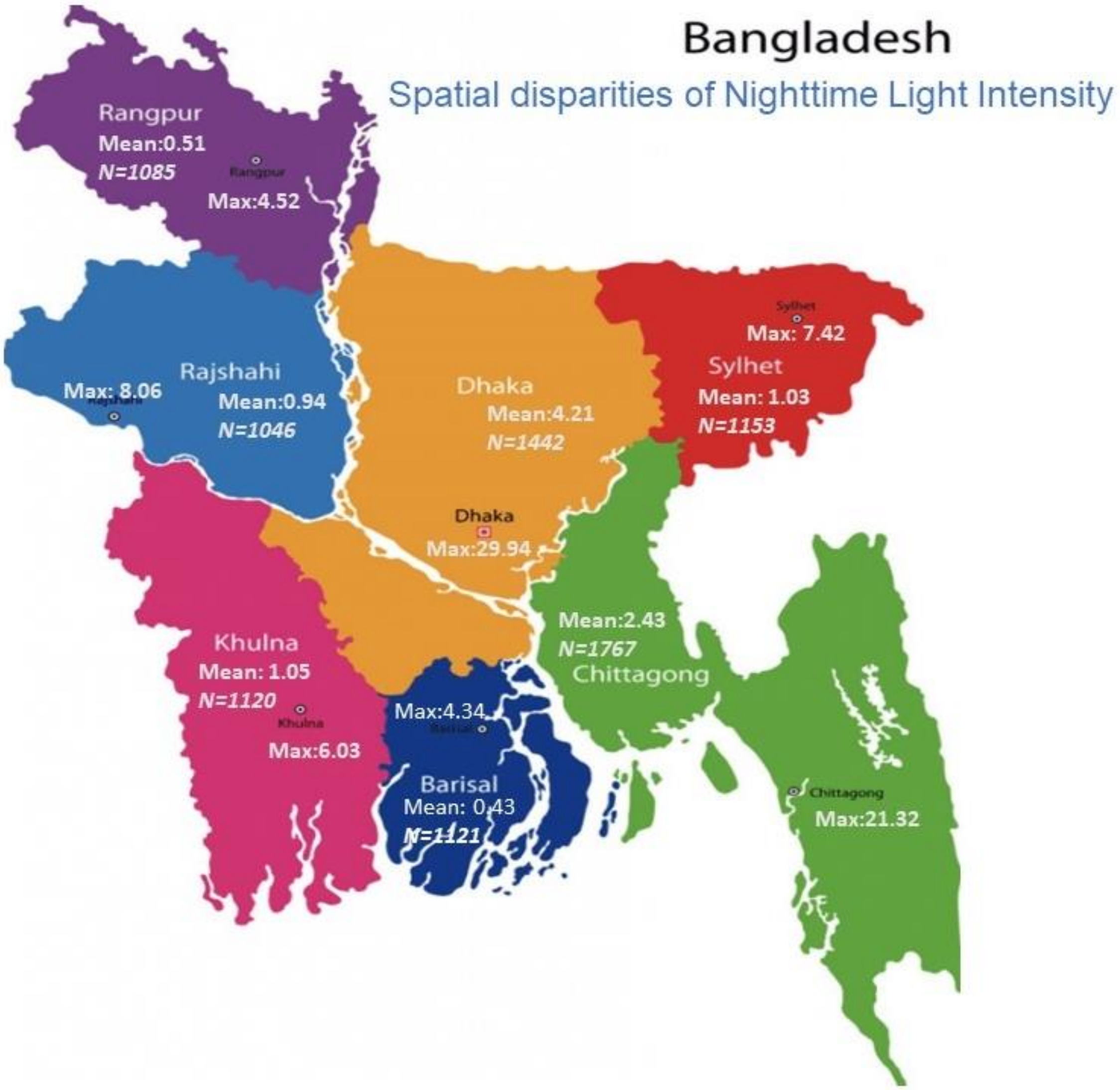

**Fig. 1**

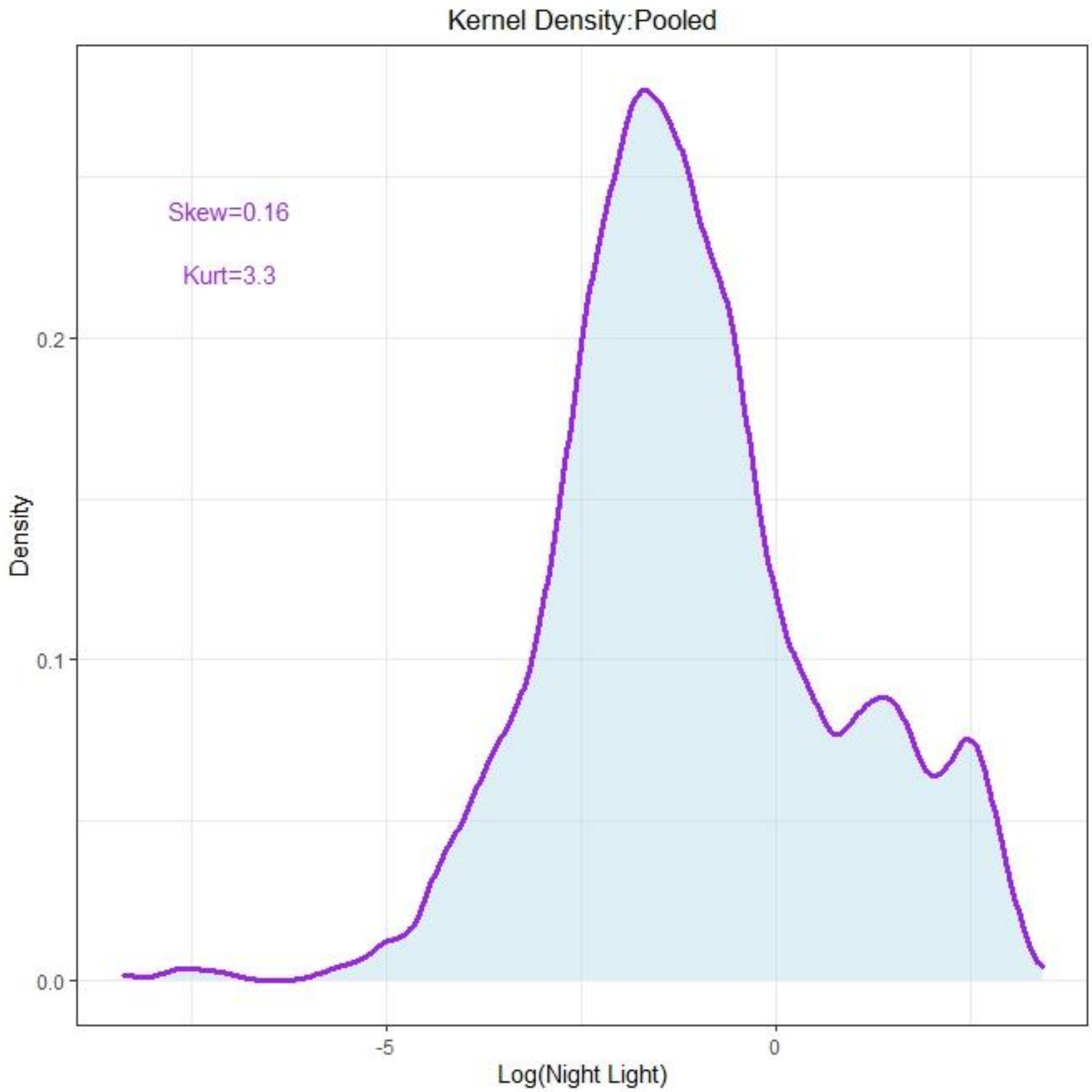

**Fig. 2a**

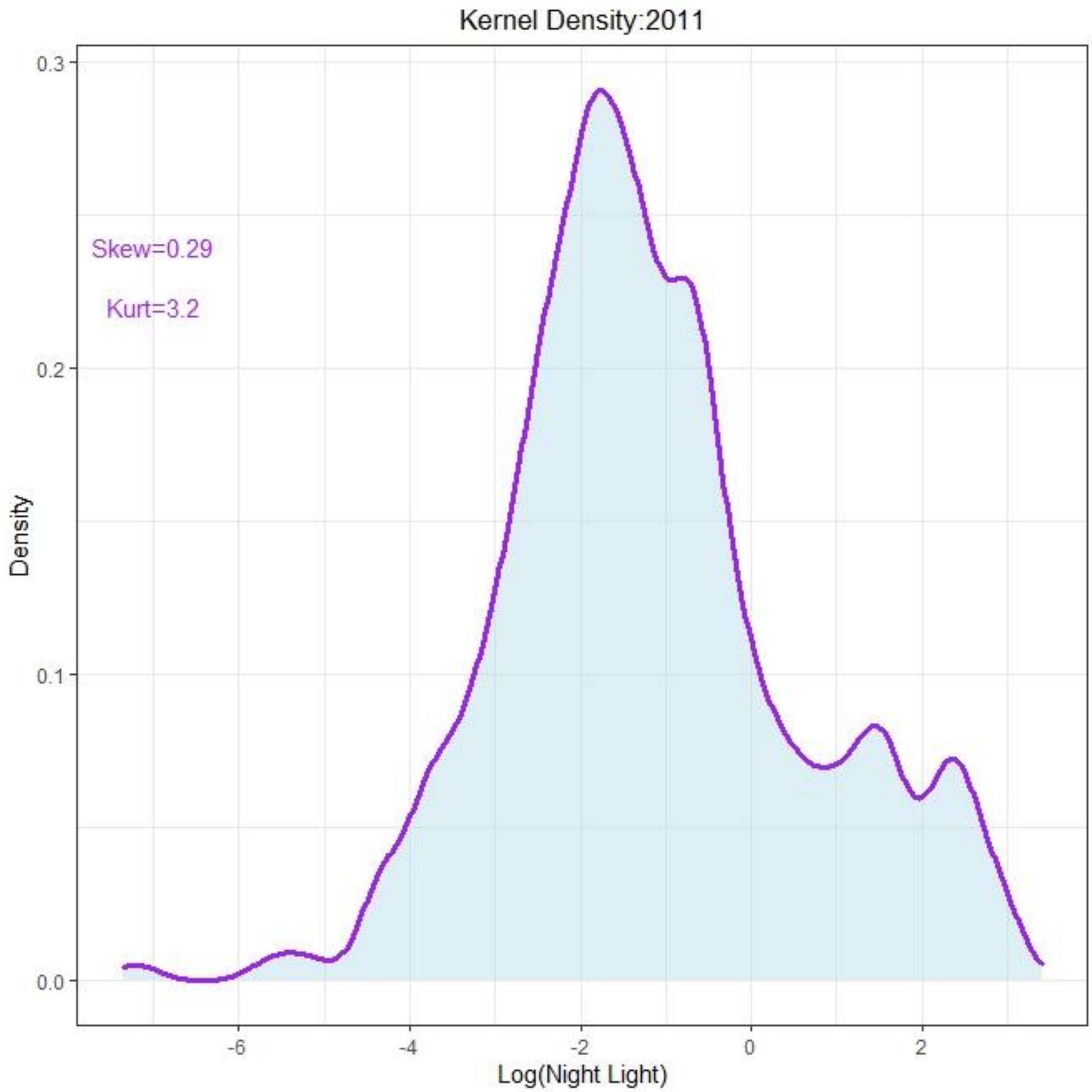

**Fig. 2b**

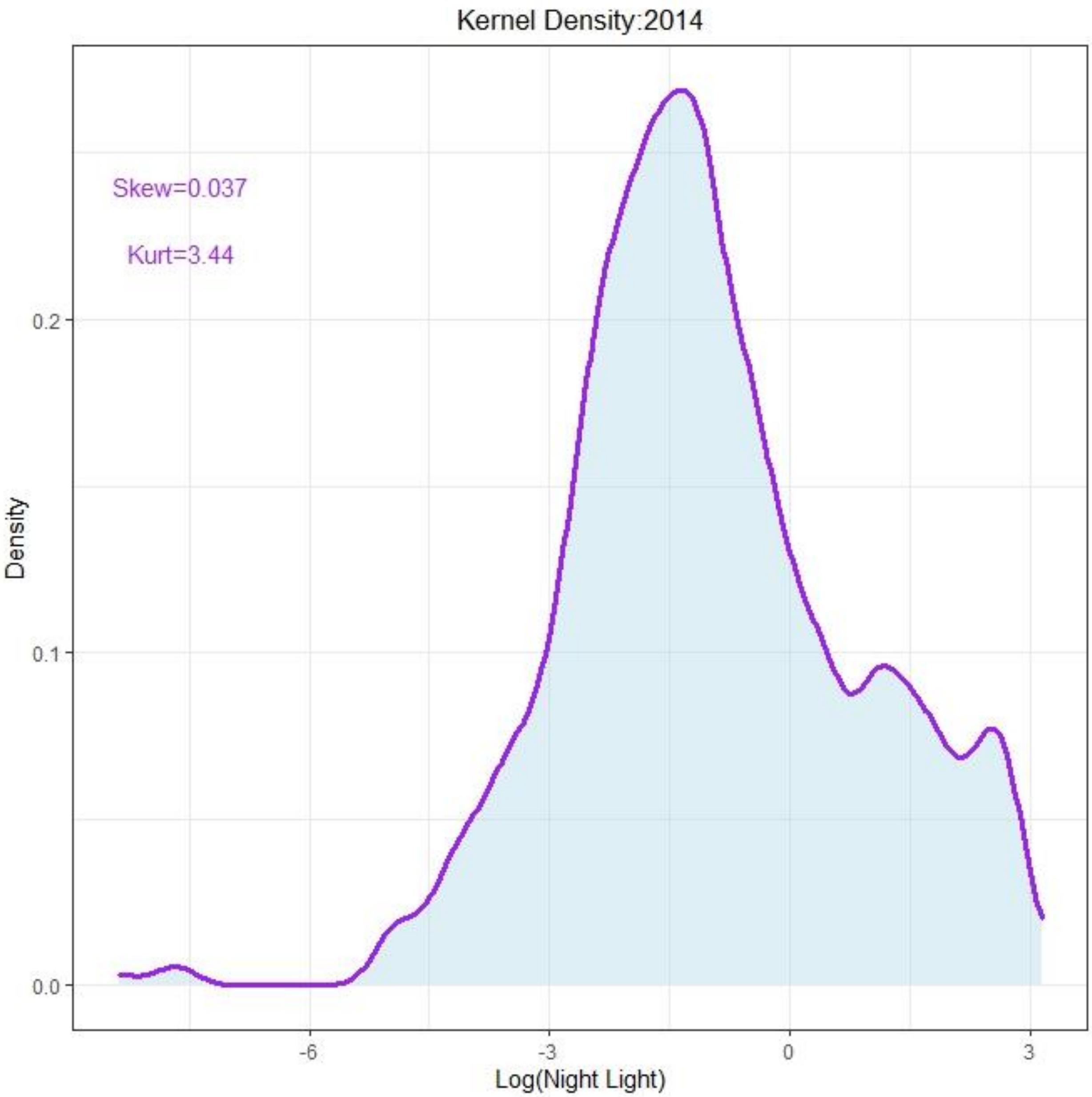

**Fig. 2c**

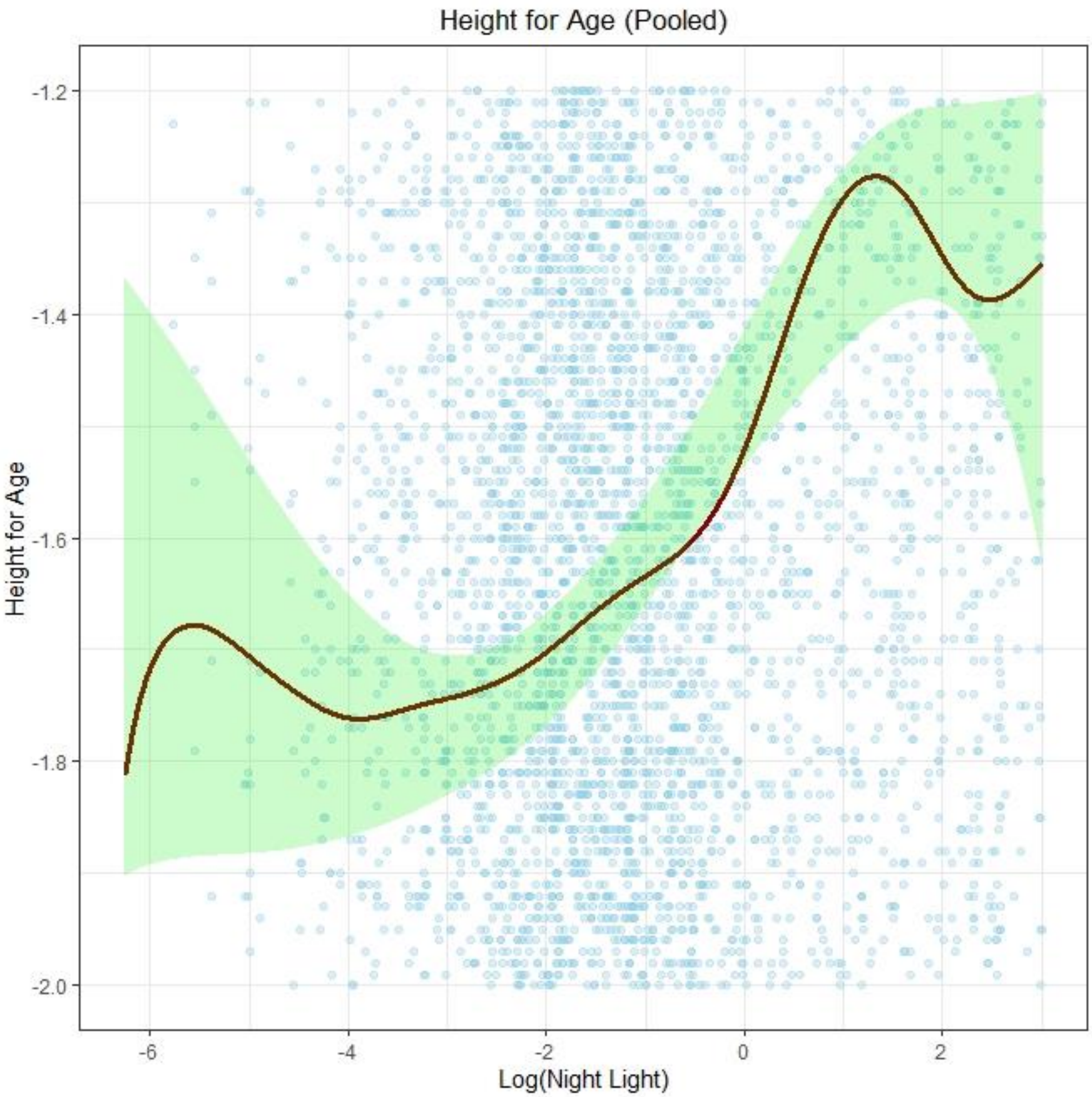

**Fig. 3a**

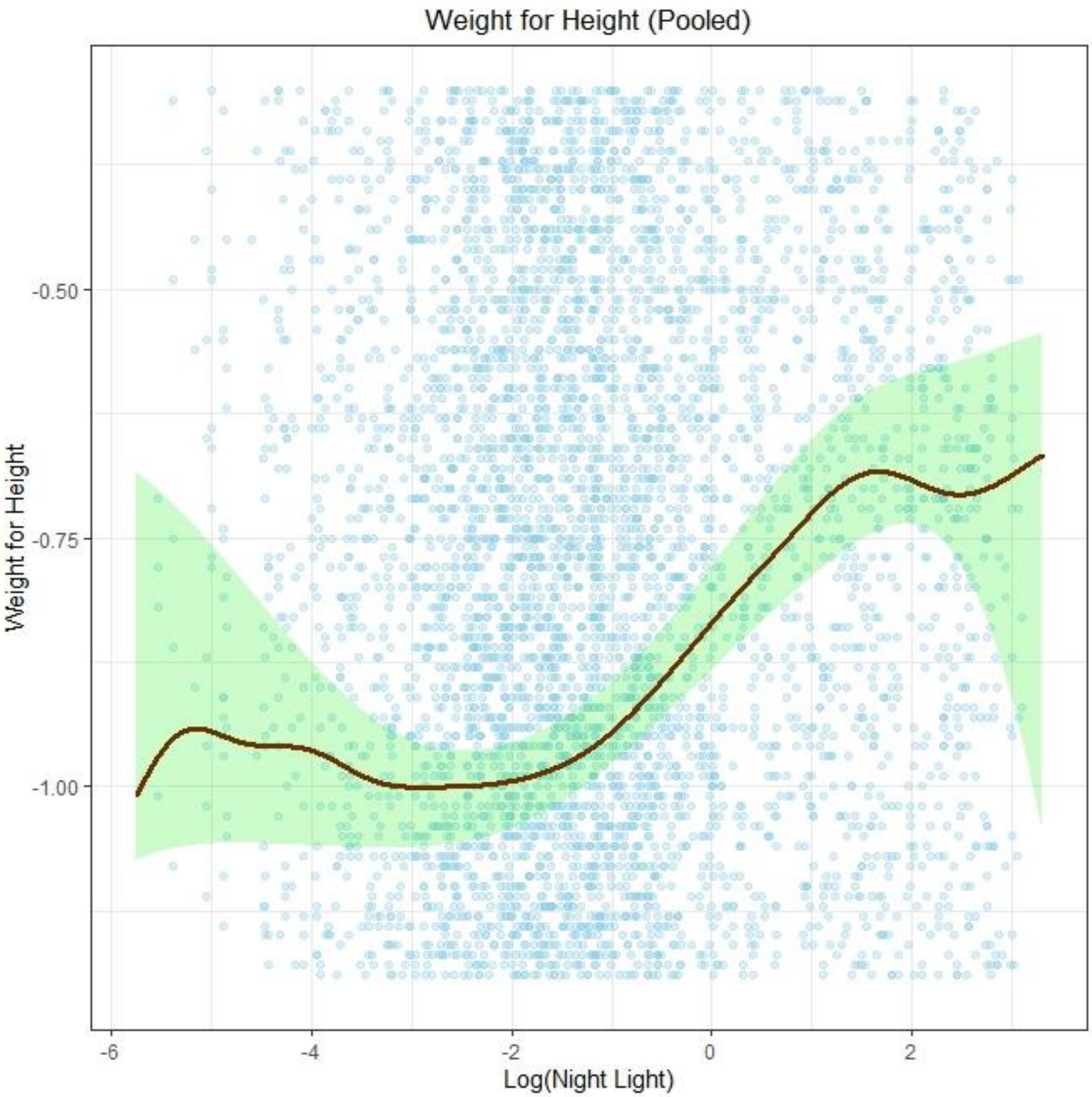

**Fig. 3b**

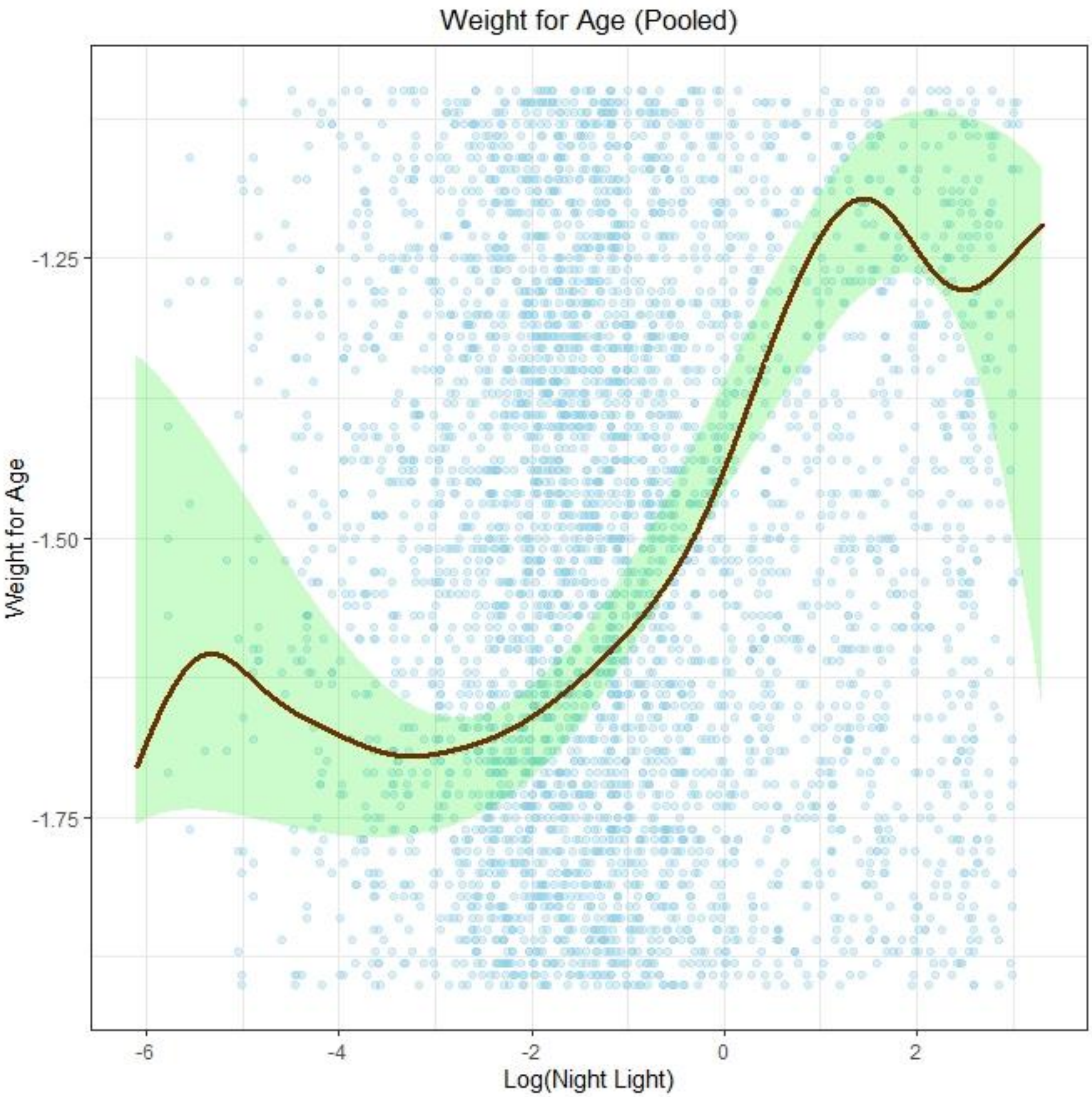

**Fig. 3c**

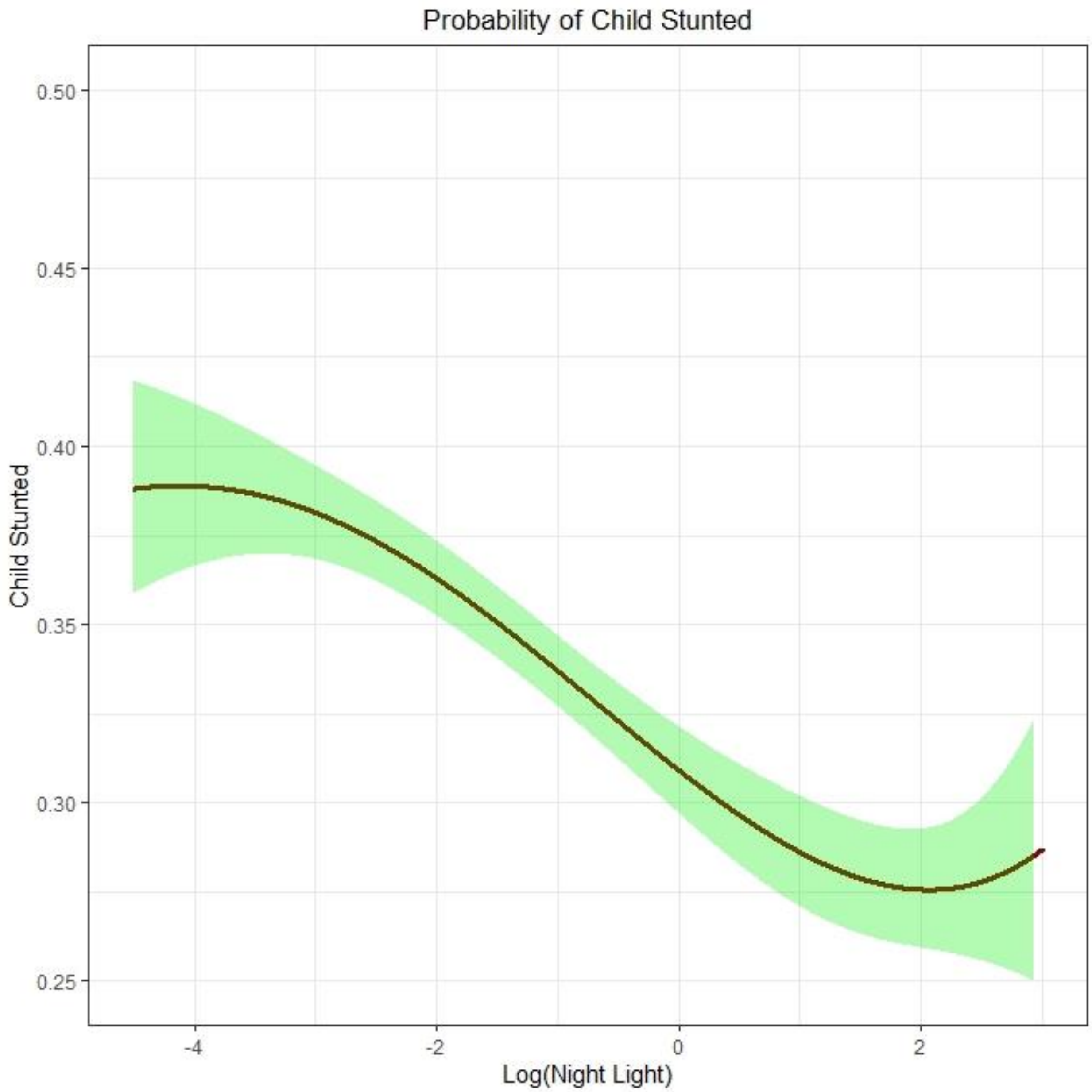

**Fig. 3d**

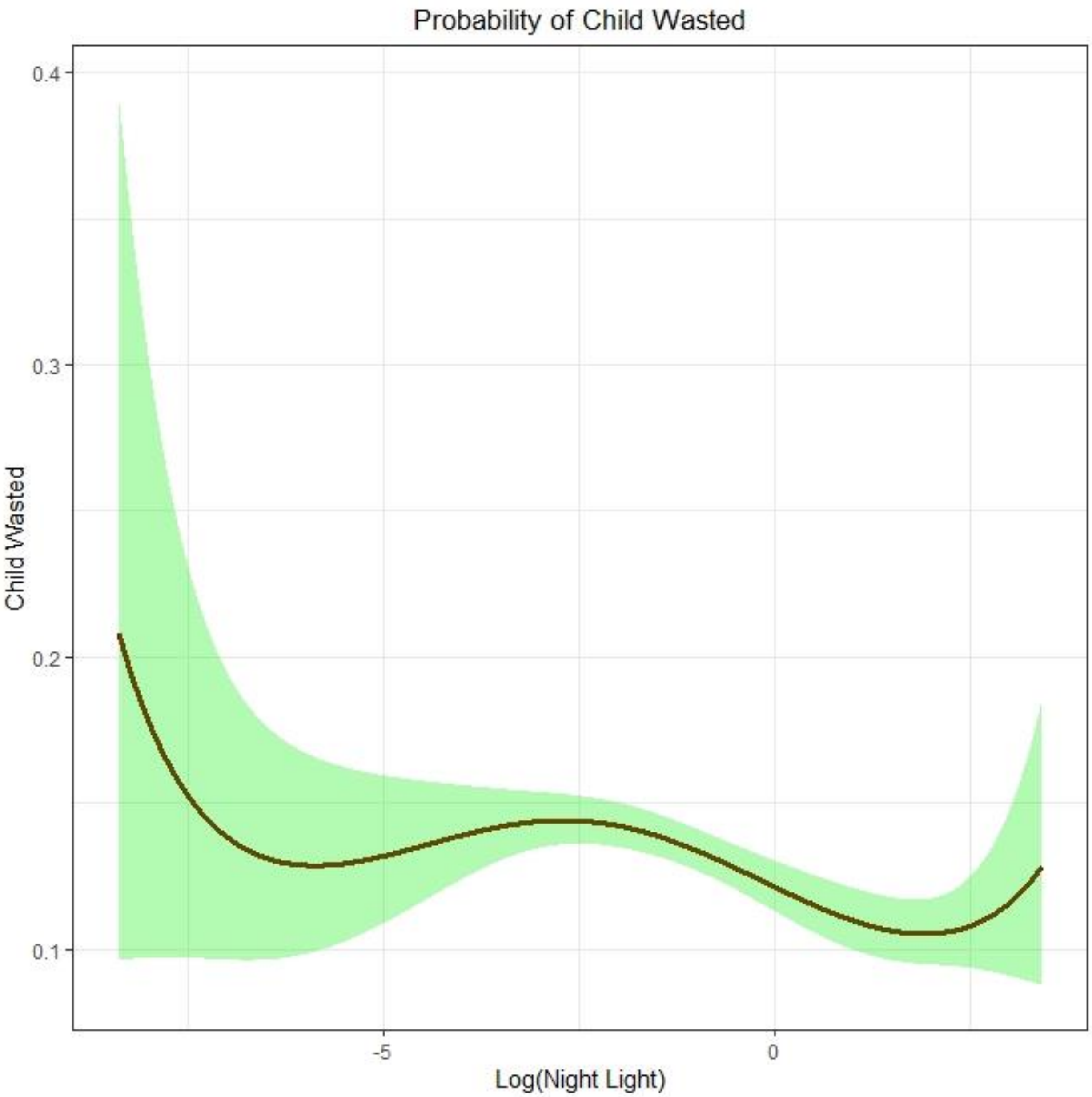

**Fig. 3e**

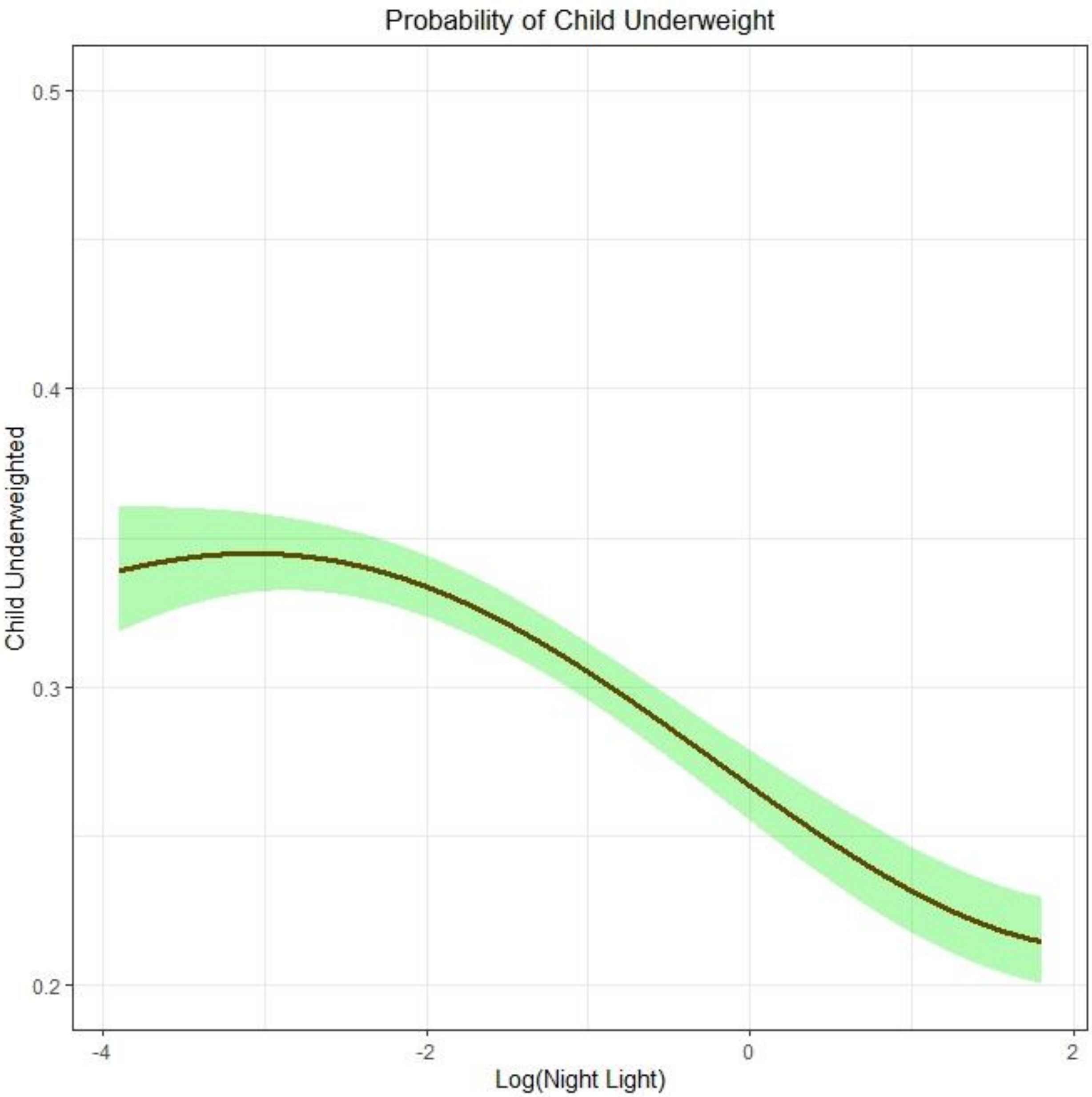

**Fig. 3f**

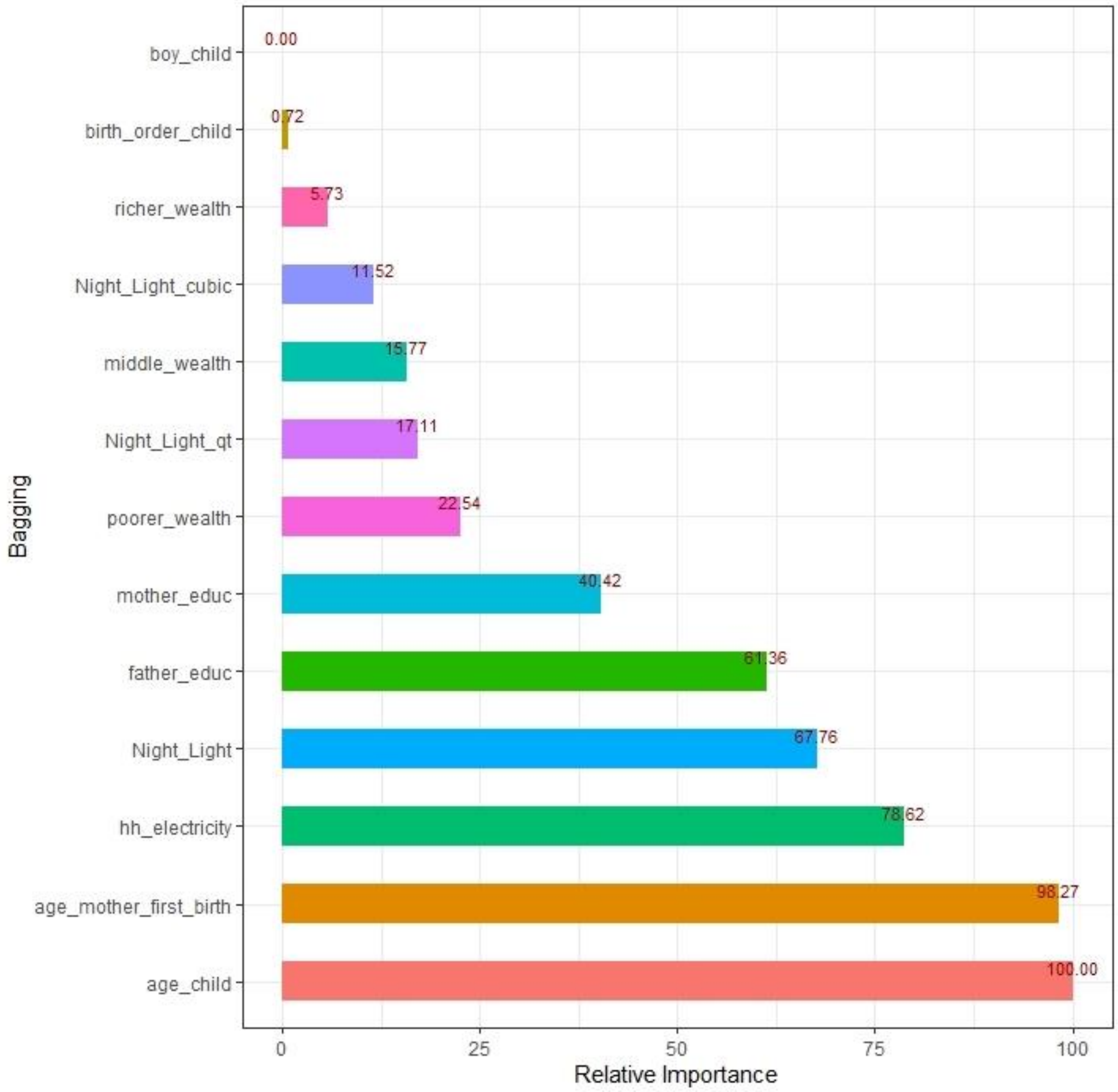

**Fig. 4a**

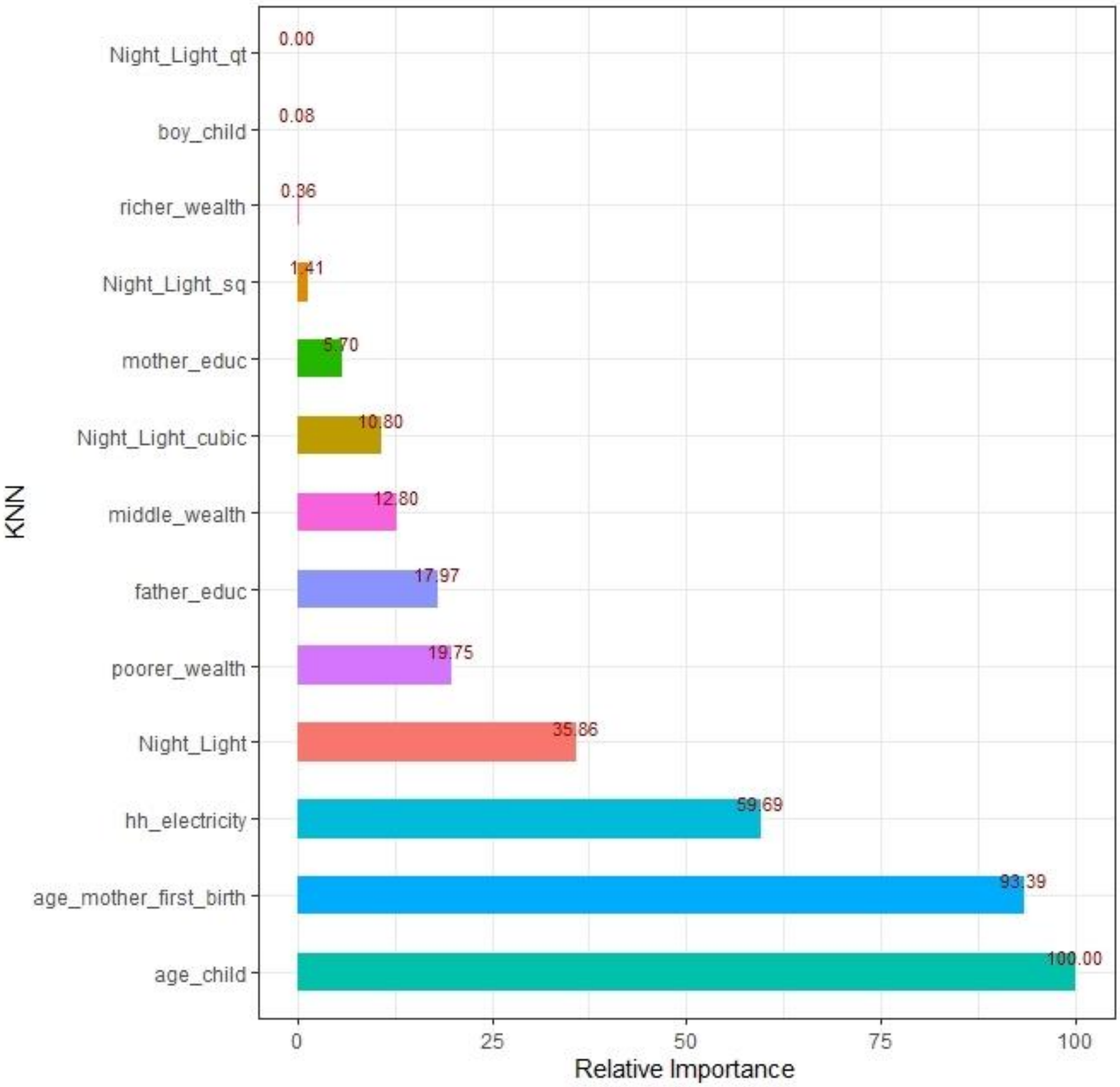

**Fig. 4b**

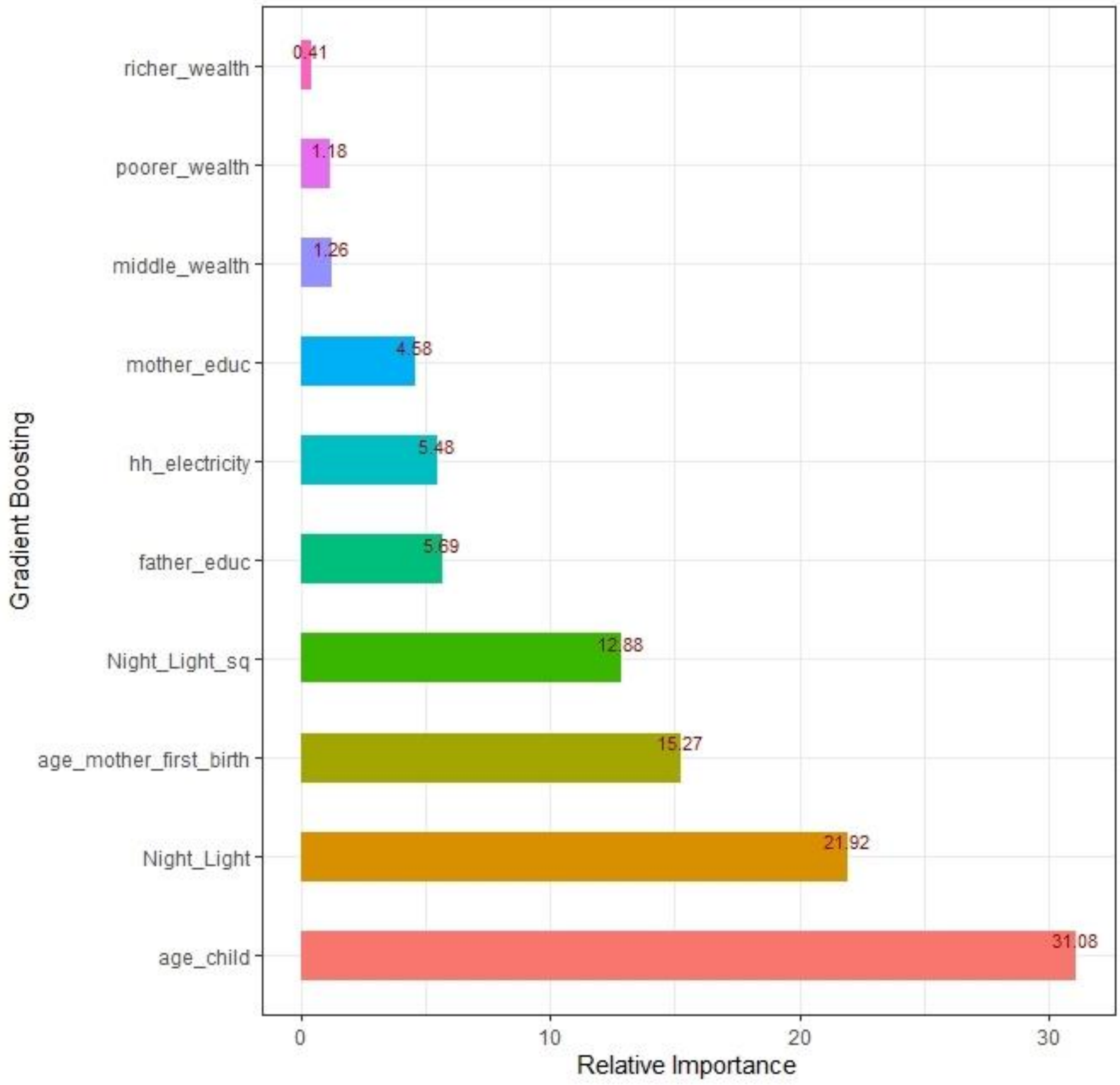

**Fig. 4c**

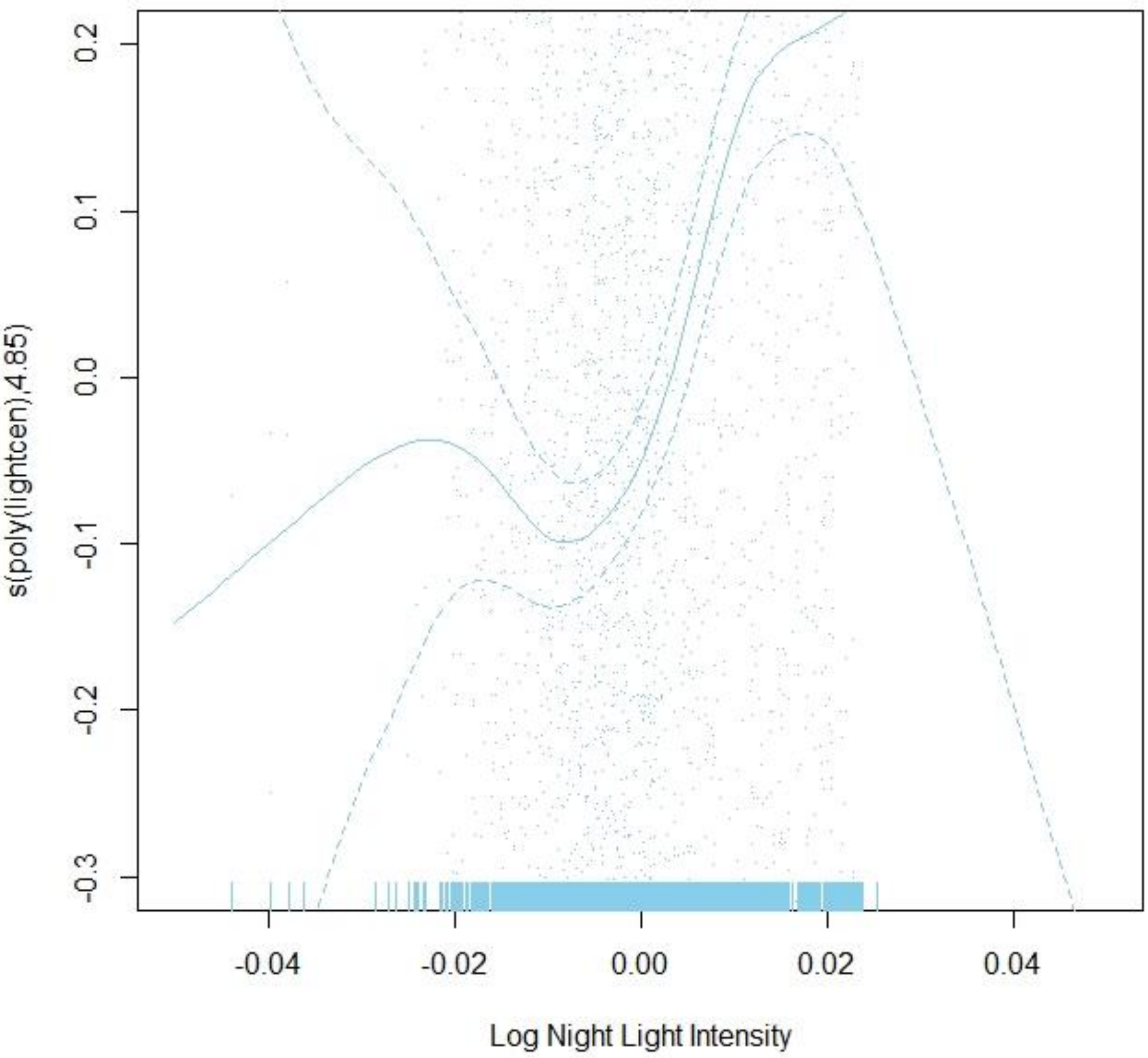

**Fig. 5**